\documentclass[journal,10pt,twocolumn]{IEEEtran}

\usepackage{cite}
\usepackage{amssymb}
\usepackage{amsmath}
\usepackage{graphicx}
\usepackage{enumitem}
\usepackage{bm}
\usepackage{mathtools}
\usepackage{amsfonts}
\usepackage{stackengine}
\usepackage{xcolor}
\usepackage{stfloats}
\usepackage{array}
\usepackage{scalerel}
\usepackage[export]{adjustbox}
\newcolumntype{M}[1]{>{\centering\arraybackslash}m{#1}}
\newcolumntype{P}[1]{>{\centering\arraybackslash}p{#1}}
\stackMath
\usepackage{multirow} % Required for multirows
\usepackage{framed}
\usepackage{fancyhdr,lipsum}
\fancypagestyle{mahmood}{%
	\fancyhf{} % clear all fields
	
	\fancyhead[L]{\scriptsize THIS WORK HAS BEEN SUBMITTED TO THE IEEE FOR POSSIBLE PUBLICATION. COPYRIGHT MAY BE TRANSFERRED WITHOUT NOTICE.}
}%
\makeatletter
\let\ps@IEEEtitlepagestyle\ps@mahmood
\makeatother

\newcommand*{\rttensorone}[1]{\bar{{#1}}}
\newcommand*{\rttensortwo}[1]{\bar{\bar{#1}}}
\newcommand{\kernel}{{\ooalign{$k$\cr\raisebox{0.2em}{\kern0.08em--}\cr}
}}
\DeclareRobustCommand{\mhl}[1]{%
	\ifmmode\text{\color{black}{$#1$}}\else{\color{black}{#1}}\fi
}
\makeatletter
\newcommand{\vast}{\bBigg@{3}}
\makeatother

\makeatletter
\@ifdefinable\@latex@chi{\let\@latex@chi\chi}
\renewcommand*\chi{{\@latex@chi\smash[t]{\mathstrut}}} % want only bottom half of \mathstrut
\makeatletter
\def\mathclap#1{\text{\hbox to 0pt{\hss$\mathsurround=0pt#1$\hss}}}
\newcommand{\raisedchi}{\raisebox{\depth}{\(\chi\)}}
\makeatother
\def\mathclap#1{\text{\hbox to 0pt{\hss$\mathsurround=0pt#1$\hss}}}

\usepackage[font=footnotesize]{caption}
\usepackage[font=footnotesize]{subcaption} % for subfigures

\usepackage{relsize}
\usepackage{bm}
\usepackage{soul}
\usepackage{empheq}
\usepackage{placeins}

\begin{document}

	\title{Distorted Wave Extended Phaseless Rytov Iterative Method for Inverse Scattering Problems}
	
\author{Amartansh~Dubey,~\IEEEmembership{Graduate Student Member,~IEEE}, and~Ross~Murch,~\IEEEmembership{Fellow,~IEEE}\thanks{This work was supported by the Hong Kong Research Grants Council with the Collaborative Research Fund C6012-20G.}\thanks{A. Dubey is with the Department of Electronic and Computer Engineering, Hong Kong University of Science and Technology (HKUST), Hong Kong, (e-mail:
		\protect{}{adubey@connect.ust.hk})}. \thanks{R. Murch is with the Department of Electronic and Computer Engineering
		and the Institute of Advanced Study both at the Hong Kong University
		of Science and Technology (HKUST), Hong Kong.}}		
	\maketitle

\begin{abstract}
In this work we present a novel linear iterative solution to an electromagnetic inverse scattering problem with phaseless data for strongly scattering, lossy media. It is based on an extended Rytov approximation that significantly widens the validity range of the conventional Rytov approximation. By modifying this extension and including it in a distorted wave iterative formulation, accurate reconstruction results for both the real and imaginary components of permittivity can be obtained. We denote the technique as the distorted wave extended phaseless Rytov iterative method (DxPRIM) and we present its derivation and numerical formulation. Using simulation and experimental examples, we demonstrate that DxPRIM can reconstruct strong, lossy scatterers with phaseless measurement data and show that it outperforms state-of-the-art phaseless techniques.
\end{abstract}
	
\IEEEpeerreviewmaketitle
	
\section{Introduction}
\label{Sec_Intro}
Inverse scattering problems (ISPs) based on the time-independent form of the wave equation (Helmholtz equation) are challenging to solve under strong scattering conditions due to their non-linearity and ill-posedness \cite{chen2018computational, 8709721, chen2020review, TGRS2}. %Strong scattering conditions arises when scattering objects have high relative permittivity $\epsilon_r = \epsilon_R+ j \epsilon_I$ and large size.
The solution of these ISPs have a wide range of important applications including microwave imaging, remote sensing, medical imaging, and non-destructive evaluation \cite{jing2018, jing2018approximate, chen2010, chen2018computational, depatla2015x, chen2020review, pastorino2010microwave, benny2020overview, DubeyTxline, Xudongchen}.

Most conventional inverse scattering techniques require full wave data consisting of both magnitude and phase measurements of the scattered field and are referred to as full data ISPs (FD-ISPs) here. However, for many applications, it is often not practical to accurately collect phase information. Measuring phase requires a high precision measurement system with accurate synchronization between multiple receive and transmit nodes. The complexity and cost of these measurement systems increases as the frequency of the probing wave increases. Furthermore, they require accurate calibration. Hence, FD-ISPs have found limited applications in practical scenarios such as large scale microwave and indoor imaging \cite{depatla2015x, dubey2021accurate, Xudongchen, DubeyTxline, chen2010subspace, deshmukh2021physics, PD_DRIM_MF}. This has given rise to the development of techniques to solve ISPs with phaseless data (denoted as PD-ISP here) \cite{chen2018computational, dubey2021accurate, Xudongchen, chen2010subspace, deshmukh2021physics}.

PD-ISPs are shown to be more ill-posed and non-linear than FD-ISPs and hence are considered more difficult to handle \cite{Xudongchen, chen2010subspace}. Even state-of-the-art non-linear methods for solving PD-ISPs \cite{chen2010, Xudongchen, chen2018computational, chen2020review}, such as subspace optimization based methods (known as PD-SOM) fail to reconstruct strong scatterers with intricate profiles. These methods can also be more prone to experimental errors and noise due to the non-linear formulation \cite{chen2010subspace, chen2018computational}. Recent deep learning based solutions to PD-ISPs \cite{Xudongchen} can moderately extend the validity range but require large datasets for training the deep learning networks. On the other hand, existing linear methods for solving PD-ISPs such as the phaseless Rytov approximation and line-of-sight radio tomography methods are computationally less expensive and robust to noise and experimental errors \cite{depatla2015x, dubey2021accurate, dubeyTGRS, DubeyTxline, PD_DRIM_MF, Patwari2017, ye2017subspace, wombell1993reconstruction}. However, there is a reduction in the validity range due to the linear approximations involved. Therefore, there is a trade-off between practicality (handling noise, experimental errors and scarcity of measurements) and validity range while using linear verses non-linear models.

In our recent work, we proposed corrections to the conventional Rytov approximation (RA) to significantly increase its validity range in strongly scattering, lossy media \cite{dubey2021accurate, dubeyTGRS}. We denoted the formulation as the extended phaseless Rytov approximation in lossy media (xPRA-LM). In terms of shape estimation, it was demonstrated to successfully reconstruct objects up to extremely high permittivity and size, which far exceeds any known techniques. Because it is a linear phaseless model, it is also robust to experimental errors and noise. However, xPRA-LM cannot directly estimate the real and imaginary parts of relative permittivity and works only for homogeneous or piecewise homogeneous scatterers.

In this work, we modify xPRA-LM and incorporate it into the distorted wave iterative framework so that we can estimate shape as well as the complex-valued permittivity of strongly scattering, lossy scatterers. The distorted wave iterative framework is a well-known iterative approach to increase the range of applicability of existing approximations including the distorted wave Born iterative method (DBIM) and distorted wave Rytov iterative method (DRIM) and other approaches derived from these \cite{ye2017subspace, PD_DRIM_MF, chen2018computational, wombell1993reconstruction}. However, only DRIM can be extended to handle phaseless data (which we denote here as PD-DRIM) \cite{PD_DRIM_MF}. PD-DRIM is shown to be useful for strong scattering conditions, but requires incident waves at multiple frequencies over a range of a few Gigahertz (requires relative bandwidth of over 100\%) in order to achieve convergence. Overall, all existing linear, non-linear and iterative PD-ISP techniques have a limited validity range. Our key contributions in this work can be summarized as follows:
\begin{enumerate}[leftmargin=0.5cm]
	\item We formulate the distorted wave method using a modification of xPRA-LM so that it is valid in strongly scattering, lossy media. We denote the technique as the distorted wave extended phaseless Rytov method (DxPRM). It linearly relates change in an arbitrary complex-valued refractive index background profile to changes in the magnitude of the total received signal power. 
	
	\item We extend DxPRM to an iterative framework which we denote as the distorted wave extended phaseless Rytov iterative method (DxPRIM) for strongly scattering, lossy media. DxPRIM iteratively solves the forward scattering problem exactly, while the inverse scattering problem is handled using the approximation DxPRM.
	
	\item Using numerical and experimental results, we show that DxPRIM achieves a significantly wider validity range and outperforms state-of-the-art PD-ISP methods in terms of accuracy when compared to benchmark scatterer profiles (such as the ``Austria" profile).
\end{enumerate}

Organization and Notation: Section \ref{Sec_prob_form} provides the formulations for the forward problem and corresponding inverse problems. The proposed DxPRM method is formulated in Section \ref{Sec_xPRA_LM} along with a brief overview of xPRA-LM \cite{dubey2021accurate}. The DxPRIM algorithm is proposed in Section \ref{Sec_DxPRIM_LM}. Simulation and experimental results are provided in Section \ref{Sec_results}. We use $\rttensortwo{{X}}$ and $\rttensorone{X}$ to respectively denote the matrix and vector forms of a discretized parameter $X$. Lower case bold letters represent position vectors and italic letters are used to represent scalar parameters.

\section{Problem Formulation}
\label{Sec_prob_form}
\begin{figure}[!h]
	\centering
	\includegraphics[width=1.8in]{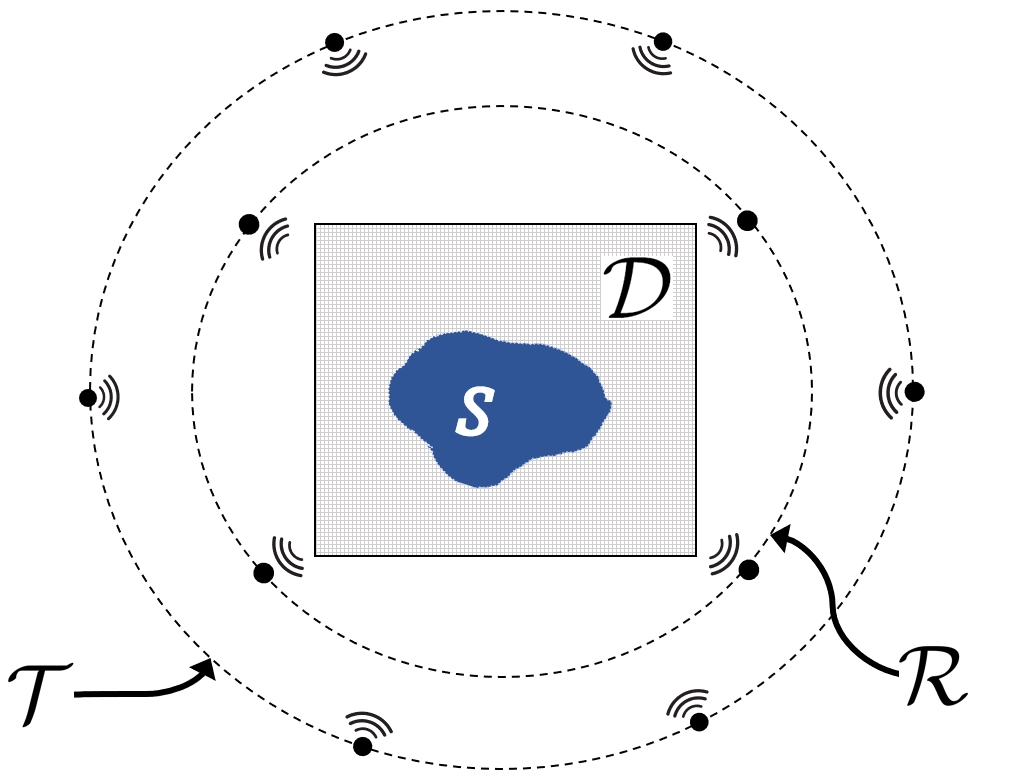}
	\caption{Illustration of the inverse scattering configuration. An object $\mathcal{{S}}$ is placed inside $\mathcal{D}$. The transmitters and receivers are placed along circular boundaries $\mathcal{T}$ and $\mathcal{R}$ respectively.}
	\label{geometry}
	\vspace{-0.2\baselineskip}
\end{figure}

Consider the setup shown in Fig. \ref{geometry} which consists of a 2D domain of interest, $\mathcal{D}$, in which there are non-magnetic scattering objects $\mathcal{S}$. The domain of interest, $\mathcal{D}$, is characterized by complex-valued relative permittivity $\epsilon_r(\bm{r}) = \epsilon_R(\bm{r})+j \epsilon_I(\bm{r})$ or equivalently by refractive index $\nu(\bm{r}) = \nu_R(\bm{r})+ j\nu_I(\bm{r}) = \sqrt{\epsilon_r(\bm{r})}$. To probe $\mathcal{D}$, an array of electromagnetic sources (or transmitters) are placed along circular boundary $\mathcal{T}$. The total number of transmitters are $M_t$ and the location of each transmitter is given by $\bm{r}_{m_t} \in \mathcal{T}$ where, $m_t = 1, 2, ..., M_t$. The radiation from the transmitters is monochromatic, time harmonic ($e^{j\omega t}$) and vertically polarized which is often referred to as transverse magnetic (TM) in inverse scattering contexts. The scattered waves from $\mathcal{D}$ are collected by an array of receivers placed along the measurement boundary $\mathcal{R}$. The total number of receivers is $M_r$ and the location of each receiver is given by $\bm{r}_{m_r} \in \mathcal{R}, m_r = 1, 2, ..., M_r$. 

Let the free-space incident field at any point inside $\mathcal{D}$ due to a transmitter (at $\bm{r}_{m_t}$) be $E_{m_t}^i(\bm{r})$. The resultant total field inside $\mathcal{D}$ (due to illumination by $E_{m_t}^i(\bm{r})$) can be written as the sum of the incident and scattered field, i.e., $E_{m_t}(\bm{r}) = E_{m_t}^i(\bm{r}) + E_{m_t}^s(\bm{r})$. The incident field $E_{m_t}^i(\bm{r})$ and total field $E_{m_t}(\bm{r})$ satisfy the homogeneous and inhomogeneous Helmholtz wave equation given by (\ref{Eq_Hzfree}a) and (\ref{Eq_Hzfree}b) respectively,
\begin{subequations}
	\label{Eq_Hzfree}
	\begin{align}
		(\nabla^2 + k_0^2) E_{m_t}^i(\bm{r}) & = 0, \qquad && \bm{r} \in \mathcal{D}\\
		(\nabla^2 + k_0^2 \nu^2(\bm{r})) E_{m_t}(\bm{r}) & = 0, \quad && \bm{r} \in \mathcal{D}
	\end{align}
\end{subequations}
where $k_0=2\pi/\lambda_0$ is the free-space wavenumber, $\lambda_0$ is the free-space wavelength and $\nu^2 = \epsilon_r$.

\subsection{Forward Scattering Problem}
The goal of the forward problem is to estimate the total field (or scattered field) at the receiver, given the permittivity profile $\epsilon_r(\bm{r})$ inside $\mathcal{D}$ and incident field $E^i$. Using (\ref{Eq_Hzfree}a) and (\ref{Eq_Hzfree}b), the forward problem can be formulated in the form of two equations. The first equation models the wave-matter interaction inside $\mathcal{D}$ to estimate the total field inside $\mathcal{D}$ (due to source at $r_{m_t}$),
\begin{equation}
	\label{Eq_VSI}
	\begin{aligned}
		E_{m_t}(\bm{r})& = E_{m_t}^i(\bm{r}) + k_0^2 \int_{\mathcal{D}} g^0(\bm{r}, \bm{r}') \chi_\epsilon(r) E_{m_t}(\bm{r}') d\bm{r}'^2 
		%& = E_{m_t}^i(\bm{r}) + k_0^2 \int_{\mathcal{D}} g^0(\bm{r}, \bm{r}') J_{m_t}(\bm{r},\bm{r}') d\bm{r}'^2,
	\end{aligned}
\end{equation}
where, $\bm{r}, \bm{r}' \in \mathcal{D}$, and $\chi_\epsilon(r) = \epsilon_r(\bm{r}')-1$ is denoted here as the contrast profile. $g^0$ is the 2D free space homogeneous Green's function defined as,
\begin{equation}
	\label{Eq_FSGF}
	\begin{aligned}
		g^0(\bm{r}, \bm{r}') = \frac{j}{4} H^{(1)}_0 (k_0|\bm{r}-\bm{r}'|).
	\end{aligned}
\end{equation}
 Equation (\ref{Eq_VSI}) is also known as the Lippmann-Schwinger equation \cite{Mittra1998, chen2018computational} and by solving it we can estimate the total field $E_{m_t}(\bm{r})$ for $\bm{r} \in \mathcal{D}$. 

Using the estimate of the total field inside $\mathcal{D}$, we can then estimate the scattered field $E_{m_t}^s(\bm{r}) = E_{m_t}(\bm{r})-E_{m_t}^i(\bm{r})$ at any given receiver (at $ \bm{r}_{m_r} \in \mathcal{R}$) using,
\begin{equation}
	\label{Eq_VSI1}
	\begin{aligned}
		E_{m_t}^s(\bm{r}_{m_r}) &= E_{m_t}(\bm{r}_{m_r})-E_{m_t}^i(\bm{r}_{m_r})\\
		&=k_0^2 \int_{\mathcal{D}} g^0(\bm{r}_{m_r}, \bm{r}') \chi_\epsilon(\bm{r}') E_{m_t}(\bm{r}') d\bm{r}'^2,
	\end{aligned}
\end{equation}
This equation describes the scattered field at the receiver as a re-radiation of the induced contrast current $J_{m_t} = \chi_\epsilon \cdot E_{m_t}$.

Together, (\ref{Eq_VSI}) and (\ref{Eq_VSI1}) describes the forward problem where (\ref{Eq_VSI}) is known as the \textit{forward state equation} and (\ref{Eq_VSI1}) is known as the \textit{forward data equation} \cite{chen2018computational}. The forward state equation can be solved using the method of moments (MoMs) where we divide $\mathcal{D}$ into $N$ discrete grids, with the position vector being the center of each grid $\bm{r}_n, n = 1,2...,N$. This discrete equation can then be formulated as a system of linear equations for all $M_t$ transmitters and $M_r$ receivers. In particular we can rewrite (\ref{Eq_VSI}) in matrix-vector form to estimate the total field (inside all $N$ grids) as
\begin{equation}
	\label{Eq_VSI_MV}
	\rttensortwo{E}^i_{\scaleto{\mathcal{D}}{3pt}} = {\rttensortwo{D}}(\epsilon_r, g^0) \cdot \rttensortwo{E}_{\scaleto{\mathcal{D}}{3pt}}
\end{equation}
where, $\rttensortwo{{E}}_{\scriptscriptstyle{\mathcal{D}}} \in \mathbb{C}^{N\times M_t}$ and $\rttensortwo{E}^i_{\scaleto{\mathcal{D}}{3pt}} \in \mathbb{C}^{N\times M_t}$ are complex-valued matrices and their $m_t^{th}$ column represents the total field and incident field inside $N$ grids due to illumination by a transmitter at $\bm{r}_{m_t}$. Matrix $\rttensortwo{{D}} \in \mathbb{C}^{N\times N}$ is known as the forward state matrix which is a function of permittivity profile $\epsilon_r(\bm{r}_n)$ in $\mathcal{D}$ and free-space Green's function $g^0(\bm{r}_{m_r}, \bm{r}_n)$ and it can be derived directly by applying MoM to (\ref{Eq_VSI}) (see detailed derivation of $\overline{\overline{D}}$ in \cite{chen2018computational, Mittra1998, 1998}).

The forward data equation in (\ref{Eq_VSI1}) can also be written in matrix-vector form using the discretization used for (\ref{Eq_VSI}) as,
\begin{equation}
	\label{Eq_VSI1_MV}
	\rttensortwo{{E}}_{\scaleto{\mathcal{R}}{3pt}} = \rttensortwo{{E}}^i_{\scaleto{\mathcal{R}}{3pt}} + \rttensortwo{{S}}(\epsilon_r, g^0) \cdot \rttensortwo{E}_{\scaleto{\mathcal{D}}{3pt}}
\end{equation}
where $\rttensortwo{{E}}_{\scaleto{\mathcal{R}}{3pt}}, \rttensortwo{{E}}^i_{\scaleto{\mathcal{R}}{3pt}} \in \mathbb{C}^{M_r\times M_t}$ are the total field and incident field matrix respectively. Their elements $[\rttensortwo{{E}}_{\scaleto{\mathcal{R}}{3pt}}]_{m_r, m_t}$ and $[\rttensortwo{{E}}^i_{\scaleto{\mathcal{R}}{3pt}}]_{m_r, m_t}$ represent the total and incident field at receiver $\bm{r}_{m_r}$, when $\mathcal{D}$ is illuminated by a transmitter at $\bm{r}_{m_t}$. Matrix $\rttensortwo{{S}}(\epsilon_r, g^0) \in \mathbb{C}^{M_r\times N}$ is the forward data problem kernel matrix which is the product of the contrast profile $\chi_\epsilon(\bm{r}')=(\epsilon_r(\bm{r}_n)-1)$ and free space Green's function $g^0(\bm{r}_{m_r}, \bm{r}_n)$, scaled by the area of one grid. Matrix $\rttensortwo{{E}}_{\scriptscriptstyle{\mathcal{D}}} \in \mathbb{C}^{N\times M_t}$ represents the total field inside $\mathcal{D}$ as estimated by solving (\ref{Eq_VSI_MV}). 

Overall, solving the forward problem (given $\epsilon_r(\bm{r}), E^i$), involves a two step process. First estimating the total field inside $\mathcal{D}$ using (\ref{Eq_VSI_MV}) and then substituting it in (\ref{Eq_VSI1_MV}) to estimate the total field at the receiver. The total field estimation using (\ref{Eq_VSI_MV}) and (\ref{Eq_VSI1_MV}) provides numerically exact results as these equations are derived from the wave equation without any approximation.

\subsection{Inverse Scattering Problem}
Data equation (\ref{Eq_VSI1}) (or (\ref{Eq_VSI1_MV})) can also be posed as an inverse problem. That is, estimation of $\chi_\epsilon$ in  $\mathcal{D}$ , given the total field $E_{m_t}(\bm{r})$ at the receivers $\bm{r}\in \mathcal{R}$. However, it is non-linear and ill-posed because along with $\chi_\epsilon$, total field $E_{m_t}(\bm{r})$ inside  $\mathcal{D}$  is also unknown in (\ref{Eq_VSI1}), which itself is a function of unknown $\chi_\epsilon$. With phaseless data, there is even less information in the data and hence (\ref{Eq_VSI1}) becomes even more severely non-linear and ill-posed.

Due to the difficulty in solving (\ref{Eq_VSI1}), approximate linear models have been proposed \cite{chen2018computational, murch1990inverse, wu2003wave}. The most common include BA and RA. These linear models have low computational complexity, and more importantly, due to a linear relation in measurements and contrast, these are less sensitive to experimental noise and errors. However, approximating a non-linear problem using linearization, results in a limited validity range. 

Despite their limited validity range, linear models have been shown to be useful when used with distorted wave iterative frameworks \cite{ye2017subspace, PD_DRIM_MF, chen2018computational}, where approximate inverse models are combined with exact forward models in an iterative framework. In distorted wave iterative frameworks (such as DBIM or DRIM), the algorithm alternates between incrementally correcting the contrast profile estimate and improving the estimate of the total field and inhomogeneous Green's function inside  $\mathcal{D}$. It updates the contrast profile and then estimates an improved estimate of the total field and inhomogeneous Green's function. This process is repeated iteratively until convergence is achieved with the best contrast profile estimate. To generate an estimate of the contrast profile at any iteration, linear models such as BA or RA are used in their distorted wave form. %, which is then substituted in forward problem state equation in (\ref{Eq_VSI_MV}) to obtain updated total field and inhomogeneous greens functions).

Of the two approximations, BA and RA, only RA can be adapted for phaseless data. The Rytov transformation normalizes the total field $E_{m_t}(\bm{r})$ by the incident field $E_{m_t}^i(\bm{r})$ to express the scattering using a complex phase $\phi_{m_t}^s(\bm{r})$,
\begin{equation}
	\label{Eq_rytTfield}
	\begin{aligned}
		\frac{E_{m_t}(\bm{r})}{E_{m_t}^i(\bm{r})} &= e^{\phi_{m_t}^s(\bm{r}) }.
		%\\
		%\implies \psi(\bm{r}) & = \psi_i(\bm{r}) e^{jk\phi_s(\bm{r})}
	\end{aligned}
\end{equation}
Using (\ref{Eq_rytTfield}), (\ref{Eq_Hzfree}b) and (\ref{Eq_Hzfree}a) we can obtain a non-linear differential equation (Riccati equation in $E_{m_t}^i \cdot \phi_{m_t}^s$) \cite{wu2003wave}, which can be written in integral form to obtain an expression for the total field at the receiver ($\bm{r}_{m_r}$) as
\begin{equation}
	\label{Eq_rytov2}
	\begin{aligned}
		&E_{m_t}\left(\bm{r}_{m_r}\right)  = \\  & E_{m_r}^i \left(\bm{r}_{m_r} \right) \cdot e^{\left(  \frac{k_0^2}{E_{m_t}^i\left(\bm{r}_{m_r}\right)} \int_{\mathcal{D}} g^0(\bm{r}_{m_r}, \bm{r'})  \chi_{\text{RI}}(\bm{r}') E_{m_t}^i(\bm{r'}) d\bm{r'}^2\right)},
	\end{aligned}
\end{equation}
where, 
\begin{equation}
	\label{Eq_rytov2_chi}
	\begin{aligned}
	\chi_{\text{RI}} & (\bm{r}) = \epsilon_r(\bm{r})-1 + \frac{\nabla \phi_{m_t}^s(\bm{r}) \cdot \nabla \phi_{m_t}^s(\bm{r})}{k_0^2}.
\end{aligned} 
\end{equation}
We refer to (\ref{Eq_rytov2}) as the Rytov integral (RI) in the remainder of this paper. Also note that $\chi_{\text{RI}}$ is the contrast function of RI and is different from the conventional definition of contrast profile $\chi_\epsilon = \epsilon_r-1$ used in the forward problem. 

The term $\nabla \phi_{m_t}^s \cdot \nabla \phi_{m_t}^s$ is neglected under a weak scattering assumption to arrive at the conventional RA formulation as,
\begin{equation}
	\label{Eq_RA}
	\begin{aligned}
		&E_{m_t}\left(\bm{r}_{m_r}\right)  = \\ & E_{m_r}^i(\bm{r}_{m_r}) \cdot e^{\left(  \frac{k_0^2}{E_{m_t}^i(\bm{r}_{m_r})} \int_{\mathcal{D}}  g^0(\bm{r}_{m_r}, \bm{r'})  \chi_{\epsilon}(\bm{r}') E_{m_t}^i(\bm{r'}) d\bm{r'}^2\right)}.
	\end{aligned}
\end{equation}
%We can easily transform the formulation of RA in (\ref{Eq_RA}) into the phaseless form by taking logarithm (base 10) both side,
%\begin{equation}
%	\label{Eq_Rytovphaseless}
%	\begin{aligned}
%		P_{m_t}& (\bm{r}_{m_r})  [\text{dB}] = P_{m_t}^i(\bm{r}_{m_r})[\text{dB}]  \ + \nonumber \\ & C_0 \cdot \operatorname{Re}\bigg(\frac{k_0^2}{E_{m_t}^i(\bm{r}_{m_r})}  \int_{\mathcal{D}} g^0(\bm{r}_{m_r}, \bm{r'})  \chi_{\text{RA}}(\bm{r'}) E_{m_t}^i(\bm{r'}) d\bm{r'}^2\bigg),
%	\end{aligned}
%\end{equation}
%where, $C_0 = 20\log_{10} e$, $P_{m_t}$ and $P^i_{m_t}$ are total power and free space incident power in dB. The operator $\operatorname*{Re}$ takes real part of its operand.

RA in (\ref{Eq_RA}) is transformed into phaseless form by multiplying (\ref{Eq_RA}) with its conjugate, (which is not possible with other approximate models such as BA),
\begin{equation}
	\label{Eq_PRA}
	\begin{aligned}
		&\left|E_{m_t}\left(\bm{r}_{m_r}\right)\right|^2  =\nonumber \\ & \left| E_{m_r}^i(\bm{r}_{m_r})\right|^2  \cdot e^{\operatorname*{Re}\left(  \frac{2 k_0^2}{E_{m_t}^i(\bm{r}_{m_r})} \int_{\mathcal{D}}  g^0(\bm{r}_{m_r}, \bm{r'})  \chi_{\epsilon}(\bm{r}') E_{m_t}^i(\bm{r'}) d\bm{r'}^2\right)},
	\end{aligned}
\end{equation}
where, $\operatorname*{Re}$ is real part operator. 
RA is useful only for weak scattering ($\epsilon_R \approx 1$) because $\nabla \phi_{m_t}^s \cdot \nabla \phi_{m_t}^s$ is not negligible when scattering is strong (even though RA does not impose a restriction on the size of the scatterer, unlike BA \cite{wu2003wave}, it fails if permittivity contrast is large). Estimation of $\nabla \phi_{m_t}^s \cdot \nabla \phi_{m_t}^s$ is difficult as it requires solving the intractable non-linear equation (\ref{Eq_rytov2}) \cite{caorsi1996rytov, nikolova2017} which has not been performed for strongly scattering, lossy media. In our recent work \cite{dubey2021accurate}, we corrected conventional RA to increase its validity range by approximately estimating $\nabla \phi_{m_t}^s \cdot \nabla \phi_{m_t}^s$ using the high frequency theory of inhomogeneous waves in strongly scattering, lossy media. This improved RA is denoted the extended phaseless Rytov approximation in lossy media (xPRA-LM). It is briefly introduced in the next section, after which we modify it and obtain its distorted wave form for distorted iterative framework.

\section{Distorted Wave Extended Phaseless Rytov Method}
\label{Sec_xPRA_LM}
In this section we first provide the background to xPRA-LM \cite{dubey2021accurate} and then propose a modification so that it can be formulated into a distorted wave form.
\subsection{xPRA-LM and its Simplification}
\begin{figure}[!h]
	\centering
	\includegraphics[width=2.2in]{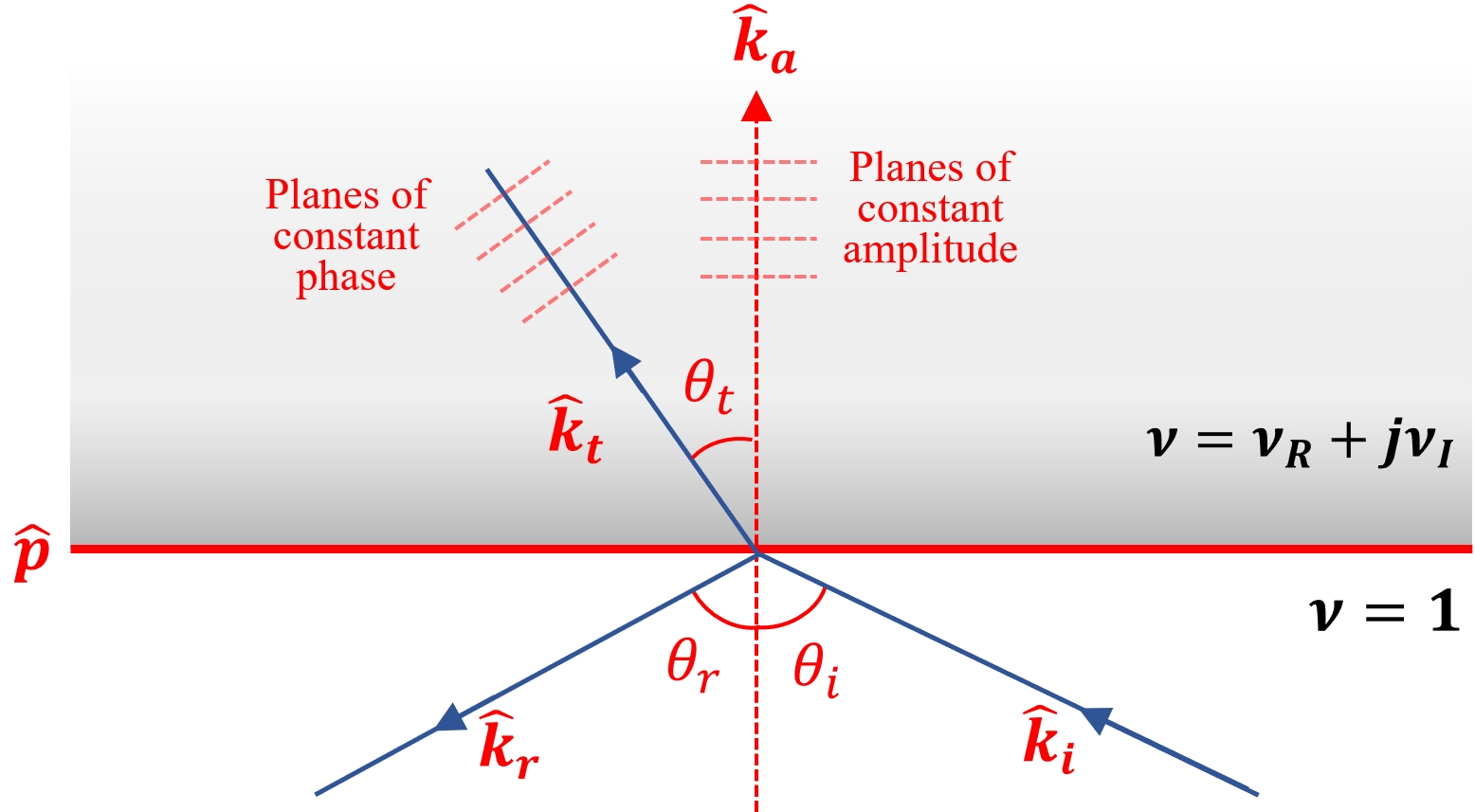}
	\caption{Free space to lossy media interface. The homogeneous plane wave (HPW) in lossless media becomes an inhomogeneous plane wave (IPW) inside the lossy media (with refractive index $\nu = \nu_R+ j \nu_I$).}
	\label{geometry-plane}
	\vspace{-0.2\baselineskip}
\end{figure}

xPRA-LM is obtained by approximately estimating the term $\nabla \phi_{m_t}^s \cdot \nabla \phi_{m_t}^s$ to correct conventional RA. To perform this we consider a plane wave in air/vacuum (in unit vector direction $\bm{\hat{k}_i}$) incident on a lossy scatterer in the upper half space as shown in Fig. \ref{geometry-plane}.

In Fig. \ref{geometry-plane} an incident plane wave traveling in the direction given by the unit vector $\bm{\hat{k}_i}$ passes into the lossy scatterer becoming an inhomogeneous plane wave (IPW) in unit vector direction $\bm{\hat{k}_t}$. The high frequency approximation \cite{dubeyTGRS, zhang2020generalized, chang2005ray} to IPW inside the scatterer can be written as
\begin{equation}
	\label{Eq_raytraced1}
	\begin{aligned}
		E_{m_t}(\bm{r}) & =A_t \ \exp{\biggl(jk_0 \int\displaylimits_{\mathclap{\text{along $\bm{\hat{k}_t}$}}} \ (V_R(\bm{r}) \bm{\hat{k}_t} +  j V_I(\bm{r}) \bm{\hat{k}_a}) \bm{dr} \biggr)},\\
	\end{aligned}
\end{equation}
where $A_t$ is transmission amplitude and $V = V_R(\bm{r}) \bm{\hat{k}_t} +  j V_I(\bm{r}) \bm{\hat{k}_a}$ is denoted the effective refractive index \cite{zhang2020generalized, chang2005ray, dubey2021accurate}.  The unit vectors $\bm{\hat{k}_t}$ and $\bm{\hat{k}_a}$ are normal to the \textit{constant phase planes} and \textit{constant amplitude planes} respectively. The concept of effective refractive index is used to mathematically decompose an inhomogeneous plane wave into a linear combination of vectors along $\bm{\hat{k}_t}$ and $\bm{\hat{k}_a}$. Equation (\ref{Eq_raytraced1}) represents a ray along path $\bm{{dr}} =dr \ \bm{\hat{k}_t}$. An interesting feature of IPW is that the planes of constant amplitude are parallel to the media interface (or scatterer boundary), in other words, vector $\bm{\hat{k}_a}$ is always normal to the media interface \cite{zhang2020generalized, chang2005ray, dubey2021accurate}. This implies that $\bm{\hat{k}_a} \cdot \bm{\hat{k}_i} = \cos\theta_i$ and $\bm{\hat{k}_a} \cdot \bm{\hat{k}_t} = \cos\theta_r$ where $\theta_i$ and $\theta_r$ are the angles of incidence and refraction respectively (see \cite{dubeyTGRS,dubey2021accurate} for more details).

Equating the total field in (\ref{Eq_raytraced1}) to (\ref{Eq_rytTfield}) and substituting the incident field as  $E_{m_t}^i(\bm{r}) = A_0 e^{j k_0 \bm{\hat{k}_i \cdot r}}$ gives
\begin{equation}
	\label{Eq_rytHFphasegrad2}
	\begin{aligned}
		&\nabla  \phi_{m_t}^s  (\bm{r}) \cdot \nabla \phi_{m_t}^s(\bm{r}) = \\
		& {k_0^2} \biggl[ V_I^2 - V_R^2 - 1 + 2 V_R (\bm{\hat{k}_t} \cdot \bm{\hat{k}_i})  - 2 j (V_R \bm{\hat{k}_t}-\bm{\hat{k}_i})  \cdot V_I \bm{\hat{k}_a} \biggr] \\ + &  (\nabla \ln T \cdot \nabla \ln T)
		+ 2 k_0 (\nabla \ln T) \biggl[j\big(V_R\ \bm{\hat{k}_t} - \bm{\hat{k}_i}\big) -   V_I \bm{\hat{k}_a} \biggr],
	\end{aligned}
\end{equation}
where $T = A_t/A_0$ is the transmission coefficient. $V_R, V_I$ are functions of $\bm{r}$ but for brevity we do not show the dependence.

Under the high frequency assumption, since $k_0$ is large, terms in (\ref{Eq_rytHFphasegrad2}) with coefficient $k_0^2$ dominate. Furthermore, the gradient of $\ln T$ is non-zero only along the boundaries of a scatterer. This makes terms containing $\nabla \ln T$ almost zero inside and outside the scatterer. Hence we can approximate $\nabla  \phi_{m_t}^s \cdot \nabla \phi_{m_t}^s$ by the dominant first term as
\begin{equation}
	\label{Eq_rytHFphasegrad3}
	\begin{aligned}
		&\nabla  \phi_{m_t}^s  (\bm{r}) \cdot \nabla \phi_{m_t}^s(\bm{r}) \approx \\
		& {k_0^2} \biggl[ V_I^2 - V_R^2 - 1 + 2 V_R (\bm{\hat{k}_t} \cdot \bm{\hat{k}_i})  - 2 j (V_R \bm{\hat{k}_t}-\bm{\hat{k}_i})  \cdot V_I \bm{\hat{k}_a} \biggr].
	\end{aligned}
\end{equation}
Under a low-loss assumption ($\delta \ll 1$) \cite{chang2005ray, zhang2020generalized}, the effective refractive index $V= V_R +  j V_I$ can be related to the actual refractive index $\nu=\nu_R + j\nu_I$ as
\begin{equation}
	\label{Eq_VRVIlowloss2}
	\begin{aligned}
		V_R \approx \nu_R, \quad 
		V_I \approx \nu_R \nu_I/\sqrt{\nu_R^2 - \sin^2\theta_i}.
	\end{aligned}
\end{equation}
Substituting $\nabla \phi_{m_t}^s \cdot \nabla \phi_{m_t}^s $ from (\ref{Eq_rytHFphasegrad3}) into (\ref{Eq_rytov2_chi}) and using (\ref{Eq_VRVIlowloss2}) provides a new approximate expression for contrast profile $\chi_{\text{RI}}$ as
\begin{equation}
	\label{Eq_rytovfulldB5}
	\begin{aligned}
		\chi_{\text{\tiny RI}}(\bm{r}) & \approx {2 (\nu_R \cos\theta_{s} -1)} + j \ { \frac{2 \nu_R \nu_I}{\sqrt{\nu_R^2 - \sin^2\theta_i}} \cos\theta_i, }\\
		& = {2 (\sqrt{\epsilon_R} \cos\theta_{s} -1)} + j \ { \frac{\epsilon_I}{\sqrt{\epsilon_R - \sin^2\theta_i}} \cos\theta_i, }
	\end{aligned}
\end{equation}
where $\cos\theta_s = \bm{\hat{k}_i} \cdot \bm{\hat{k}_t}  $ is the scattering angle (angle between incident and refracted rays) and the refractive index $\nu = \nu_R+j\nu_I$ is related to relative permittivity $\epsilon_r = \epsilon_R+j\epsilon_I$ as $\nu = \nu_R+j\nu_I = \sqrt{\epsilon_R} + j \frac{\epsilon_I}{2 \sqrt{\epsilon_R}}$ under the low-loss assumption. Note that there is a fundamental difference between the real and imaginary parts of the contrast derived in (\ref{Eq_rytovfulldB5}) compared to conventional RA ($\chi_\text{RA} = \epsilon_r(\bm{r}) -1 = \nu^2(\bm{r})-1$). Unlike RA where contrast profile is a linear function of permittivity, now the new approximate contrast profile is a non-linear function of permittivity. %In (\ref{Eq_rytovfulldB5}), the imaginary part of contrast $\operatorname*{Im}({\chi_{\text{RI}}})$ is a function of incident angle $\theta_i$ and both the real and imaginary parts of permittivity ($\epsilon_R, \epsilon_I$). While the real part $\operatorname*{Re}({\chi_{\text{RI}}})$ is a function of the real part of the permittivity $\epsilon_R$ and the scattering angle $\theta_s$ which further depends on $\theta_i$ and permittivity. The dependence of $\operatorname*{Im}({\chi_{\text{RI}}})$ on $\theta_i$ instead of $\theta_s$ is the most important aspect of the derived result. The incident angle $\theta_i$ does not change with the permittivity of the object. This implies that any distortion in the imaginary component, $\operatorname*{Im}({\chi_{\text{RI}}})$, of the reconstruction due to the presence of the $\theta_i$ terms is independent of the objects permittivity. Therefore, if the imaginary component of the reconstruction is accurate at low permittivity, it is likely to be accurate for all permittivity levels.
Also, the proposed contrast in (\ref{Eq_rytovfulldB5}) is also valid under strong scattering as the term $\nabla \phi_s \cdot \nabla \phi_s$ is not neglected unlike in RA. 
To use the approximate contrast profile (\ref{Eq_rytovfulldB5}) in a distorted wave form (then later in a iterative framework), it needs to be further simplified. The most straightforward simplification is to remove  $\theta_s$ and $\theta_i$ dependence by approximating them as $\theta_s=0$ and $\theta_i=0$. This approximation will be valid if the waves are predominantly close to normal incidence. This gives our approximation as
\begin{equation}
	\label{Eq_rytovfulldB8}
	\begin{aligned}
		\chi_{\text{\tiny RI}}(\bm{r})  & \approx 2(\nu_R-1) + j 2\nu_I\\
		 &= 2(\nu-1)\\ &= 2(\sqrt{\epsilon_r}-1) && \ \ (\text{using } \epsilon_r =\epsilon_R+j\epsilon_I = \nu^2)\\
		 &= \chi_{\text{\tiny xRA}}
	\end{aligned}
\end{equation}
where the approximation to $\chi_{\text{RI}}$ is denoted $\chi_{\text{xRA}}$. This approximation will produce estimation error, but using the distorted wave iterative framework will hopefully compensate for it.  

It can be seen that (\ref{Eq_rytovfulldB8}) is linearly proportional to the refractive index. This agrees with Fermat's principle where the incremental phase change of a ray is directly related to the product of the path length along the ray and refractive index contrast $(\nu(\bm{r}) -1)$. In other words, the incremental phase change of a ray per wavelength should be proportional to $k_0(\nu-1)$. For conventional RA, it is known (using asymptotic techniques) \cite{bates1976extended, murch1990inverse} that the incremental phase change per wavelength is $\frac{1}{2} k_0 (\nu^2 -1)$ which does not match the expected phase change as per Fermat's principal. Therefore, xPRA-LM with modified contrast in (\ref{Eq_rytovfulldB8}) also appears to better satisfy the underlying physics of the problem.

We can substitute (\ref{Eq_rytovfulldB8}) in RI (\ref{Eq_rytov2}) to obtain a phaseless form (similar to (\ref{Eq_PRA})) by multiplying (\ref{Eq_rytov2}) with its conjugate. In the resultant phaseless equation, we transform to received signal strength (RSS) in dB and new contrast (in (\ref{Eq_rytovfulldB8})) to obtain modified xPRA-LM as,
\begin{equation}
	\label{Eq_xPRA}
	\begin{aligned}
		& P_{m_t} (\bm{r}_{m_r})  [\text{dB}] - {P}_{m_t}^i(\bm{r}_{m_r})[\text{dB}]  = 20 \log_{10} \left|\frac{E_{m_t}\left(\bm{r}_{m_r}\right) }{E_{m_t}^i\left(\bm{r}_{m_r}\right) }\right| \\ & = C_0 \cdot \operatorname{Re}\bigg(\frac{k^2}{E_{m_t}^i(\bm{r})}  \int_{D} g^0(\bm{r}_{m_r}, \bm{r'}) \chi_\text{\tiny xRA}(\bm{r'}) E_{m_t}^i(\bm{r'}) d\bm{r'}^2\bigg),
	\end{aligned}
\end{equation}
where, $C_0 = 20\log_{10} e$, $P_{m_t}$ and $P^i_{m_t}$ are total power and free space incident power in dB. %The operator $\operatorname*{Re}$ takes real part of its operand

\subsection{DxPRM Formulation}
Using the contrast derived in (\ref{Eq_rytovfulldB8}), we can now formulate its use in a distorted wave form DxPRM. We estimate perturbation $\Delta \chi_\text{\tiny xRA}$ given that the refractive index and relative permittivity distribution before the perturbation are $\nu^b(\bm{r})$ and $\epsilon_r^b(\bm{r}) = (\nu^b(\bm{r}))^2$ respectively. The background incident field $E_{m_t}^b$, incident power $P_{m_t}^b (\propto 20\log(|E^b|))$ and inhomogeneous Green's function $g^b$ for this inhomogeneous background are given (or estimated by solving the forward problem for $\epsilon_r^b(\bm{r})$). Letting the refractive index, relative permittivity and total received power after the perturbation be $\nu(\bm{r})$, $\epsilon_r(\bm{r}) = \nu^2(\bm{r})$ and $P_{m_t}$ respectively, DxPRM can be written as,
\begin{equation}
	\label{Eq_DW_xPRA}
	\begin{aligned}
		& P_{m_t}  (\bm{r}_{m_r})  [\text{dB}] - {P}_{m_t}^b(\bm{r}_{m_r})[\text{dB}]  \ = \\ & C_0 \cdot \operatorname{Re}\bigg(\frac{k_0^2}{E_{m_t}^b(\bm{r}_{m_r})}  \int_{D} g^b(\bm{r}_{m_r}, \bm{r'})  \Delta \raisedchi_\text{\tiny xRA}(\bm{r'}) E_{m_t}^b(\bm{r'}) d\bm{r'}^2\bigg),
	\end{aligned}
\end{equation}
where $\Delta \raisedchi_\text{\tiny xRA} = 2(\nu-\nu^b) = \Delta \nu$ is the change in contrast and is linearly proportional to the change in refractive index. The relative permittivity profile after perturbation can be given as,
\begin{equation}
	\label{Eq_DW_xPRA_epr}
	\begin{aligned}
		\epsilon_r = \nu^2 = \left(\frac{\Delta \raisedchi_\text{\tiny xRA}}{2} + \sqrt{\epsilon_r^b}\right)^2
	\end{aligned}
\end{equation}
Equation (\ref{Eq_DW_xPRA}) can be solved using MoM where we divide $\mathcal{D}$ into $N$ discrete grids and then extend the discretized equation for all $M = M_t\cdot M_r$ transmitter-receiver links, resulting in a system of linear equations,
\begin{equation}
	\label{Eq_DW_xPRA_MoM}
	\rttensorone{P} = \rttensorone{P}^b + \operatorname*{Re}\left(\rttensortwo{\text{H}}^b(g^b, E^b) \cdot {\Delta \rttensorone{\raisedchi}_\text{\tiny xRA}}\right)
\end{equation}
where, $\Delta \rttensorone{\raisedchi}_\text{\tiny xRA} \in \mathbb{C}^{N\times 1}$ is the contrast increment vector.  $\rttensorone{P} \in \mathbb{R}^{M \times 1}$ and $\rttensorone{P^b} \in \mathbb{R}^{M \times 1}$ are total power (at receivers) after and before the perturbation respectively. $\rttensortwo{H}^b \in \mathbb{C}^{M \times N}$ is DxPRM kernel matrix which is a function of the inhomogeneous background Green's function $g^b$ and background total field $E^b$. The elements $[\rttensortwo{\text{H}}^b]_{m,n}$ are given as,
\begin{equation}
	\label{Eq_DW_xPRA_H}
	[\rttensortwo{\text{H}}^b]_{m,n} = C_0 \cdot \bigg(\frac{k_0^2}{E_{m_t}^b(\bm{r}_{m_r})} g^b(\bm{r}_{m_r}, \bm{r}_n) E_{m_t}^b(\bm{r}_n) \Delta a\bigg).
\end{equation}
The operator $\operatorname*{Re}$ can be removed by rewriting (\ref{Eq_DW_xPRA_MoM}) as,
\begin{equation}
\label{Eq_discrete3}
\rttensorone{P} - \rttensorone{P}^b = \left[ \operatorname{Re}({\rttensortwo{\text{H}}^b}) \ \ \ -\operatorname{Im}({\rttensortwo{\text{H}}^b}) \ \right] \begin{bmatrix}
	\operatorname{Re}({\Delta \rttensorone{\raisedchi}_\text{\tiny xRA}}) \\
	\operatorname{Im}({\Delta \rttensorone{\raisedchi}_\text{\tiny xRA}})
\end{bmatrix} 
\end{equation}
which can be written in compact form by substituting $\rttensortwo{\mathcal{{H}}}^b = \big[ \operatorname{Re}({\rttensortwo{\text{H}}^b}) \ \ \ -\operatorname{Im}({\rttensortwo{\text{H}}^b}) \ \big] \in \mathbb{R}^{M \times 2N}$,
\begin{equation}
	\label{Eq_discrete8}
	\Delta \rttensorone{P}^b = \rttensortwo{{\mathcal{{H}}}}^b(g^b, E^b)  
	\begin{bmatrix}
		{\Delta \rttensorone{\raisedchi}_{\tiny R}} \\
		{\Delta \rttensorone{\raisedchi}_{\tiny I}}
	\end{bmatrix}
\end{equation}
where, $\Delta \rttensorone{P^b} = \rttensorone{P}-\rttensorone{P^b}$ is change in RSS value (in dB) due to perturbation in the given background medium (with permittivity $\epsilon_r^b(\bm{r})$). The vectors $\Delta \rttensorone{\raisedchi}_{\tiny R}$ and $\Delta \rttensorone{\raisedchi}_{\tiny I}$ represent $\operatorname{Re}({\Delta \rttensorone{\raisedchi}_\text{\tiny xRA}})$ and $\operatorname{Im}({\Delta \rttensorone{\raisedchi}_\text{\tiny xRA}})$ respectively. 

When the background is free-space ($\epsilon_r=1, \Delta \raisedchi_{\text{\tiny xRA}} = \raisedchi_{\text{\tiny xRA}},  E^b = E^i, P^b = P^i, g^b = g^0$), DxPRM in (\ref{Eq_discrete8}) reduces to the discrete form of modified xPRA-LM (\ref{Eq_xPRA}) to provide a linear system of equations,
\begin{equation}
	\label{Eq_discrete9}
	\Delta \rttensorone{P} = \rttensortwo{{\mathcal{{H}}}}^0(g^0, E^i)  
	\begin{bmatrix}
		{ \rttensorone{\raisedchi}_{\tiny R}} \\
		{ \rttensorone{\raisedchi}_{\tiny I}}
	\end{bmatrix}.
\end{equation}

The inverse problem utilizing (\ref{Eq_discrete8}) and (\ref{Eq_discrete9}) is ill-posed as there are usually significantly fewer measurements $(M = M_t \cdot M_r )$ compared to the number of grids $(N)$ with unknown permittivity, i.e., $M\ll N$. Furthermore, due to the removal of $\operatorname*{Re}$ operator, we need to estimate $[	{\Delta \rttensorone{\raisedchi}_{\tiny R}} \ \
{\Delta \rttensorone{\raisedchi}_{\tiny I}}]^T$ which is a $2N \times 1$ vector, making the problem more severely ill-posed. Hence, $\rttensortwo{\mathcal{H}}$ or its pseudo inverse is very poorly conditioned. To tackle this, regularization is needed. We use Tikhonov regularization to rewrite the optimization objective function for DxPRM (\ref{Eq_discrete8}) as,
\begin{equation}
	\label{Eq_Regularize8}
	\begin{aligned}
		\underset{\tiny {\Delta \rttensorone{\raisedchi}_{\tiny R}, \Delta \rttensorone{\raisedchi}_{\tiny I} }}{\text{min}} \ \ \ \left|\left|\Delta \rttensorone{P}^b-\rttensortwo{\mathcal{H}}^b	\begin{bmatrix}
			{\Delta \rttensorone{\raisedchi}_{\tiny R}} \\
			{\Delta \rttensorone{\raisedchi}_{\tiny I}}
		\end{bmatrix} \right|\right|^2_2 + {\beta} \ \left|\left| \Gamma
		 \begin{bmatrix}
		{\Delta \rttensorone{\raisedchi}_{\tiny R}} \\
		{\Delta \rttensorone{\raisedchi}_{\tiny I}} 
	\end{bmatrix} \right|\right|^2_2
	\end{aligned}
\end{equation}
where $\Gamma \in \mathbb{C}^{2N\times 2N} $ is the Tikhonov matrix which imposes the required prior on the optimization variable. The straightforward choice for $\Gamma$ is the identity matrix which reduces  (\ref{Eq_Regularize8}) to $l_2$ regularization (or Ridge regression). $\Gamma$ can also be the finite difference matrix to impose a smoothness prior. The smoothness prior is known to be better than Ridge in removing noise. However, Ridge requires less computation because the smoothness prior requires two finite difference matrices to be used (to impose smoothness in horizontal and vertical direction). In this work, we use Ridge for reconstructing homogeneous scatterers whereas we can use the smoothness prior for more intricate inhomogeneous scatterers (note both can achieve the same final results and only the number of iterations required to converge to the optimum solution differs). 

%It is important to note that an advantage of using a straightforward regularization such as Tikhonov (with identity Tikhonov matrix) is that an objective of the proposed iterative framework (DxPRIM) is also to learn sparsity iteratively (by removing unwanted distortions due to approximate inverse model).
It is important to note that using the straightforward Tikhonov prior (with identity Tikhonov matrix) instead of an advanced sparsity and structure promoting prior such as  TVAL3 \cite{li2009user} can help us make sure that the sparsity and distortion removal is being learned by the proposed DxPRIM framework instead of the advanced regularization techniques.

The analytical solution for DxPRM (with Tikhonov regularization (\ref{Eq_Regularize8})) can be written as,
\begin{equation}
	\label{Eq_DW_xPRA9_reg}
	\begin{aligned}
		\begin{bmatrix}
			{\Delta \rttensorone{\raisedchi}_{\tiny R}} \\
			{\Delta \rttensorone{\raisedchi}_{\tiny I}} 
		\end{bmatrix} = \left(\big({\rttensortwo{\mathcal{H}}^b}\big)^T \rttensortwo{\mathcal{H}}^b + \beta \Gamma^T \Gamma \right)^{-1} \big({\rttensortwo{\mathcal{H}}^b}\big)^T \Delta \rttensorone{P}^b
	\end{aligned}
\end{equation}
where $(\cdot)^T$ is the transpose of matrix (or Hermitian of matrix for complex-valued matrix). Similarly, the analytical solution when background is free-space can be written as,
\begin{equation}
	\label{Eq_DW_xPRA8_reg}
	\begin{aligned}
		\begin{bmatrix}
			{ \rttensorone{\raisedchi}_{\tiny R}} \\
			{ \rttensorone{\raisedchi}_{\tiny I}} 
		\end{bmatrix} = \left(\big({\rttensortwo{\mathcal{H}}^0}\big)^T \rttensortwo{\mathcal{H}}^0 + \beta \Gamma^T \Gamma \right)^{-1} \big({\rttensortwo{\mathcal{H}}^0}\big)^T \Delta \rttensorone{{P}}
	\end{aligned}
\end{equation}
and we refer to this as the modified xPRA-LM. 

% We can also look at (\ref{Eq_rytovfulldB5}) from the perspective of Fermat's principle to gain more insight \cite{bates1976extended}. Under strong scattering  ($\epsilon_R\gg1$), and for the special cases of normal incidence ($\theta_i=0$) and forward scattering ($\theta_s=0$), our result  (\ref{Eq_rytovfulldB6-1}) reduces to refractive index $\chi_{\text{RI}} \approx 2 (\nu_R-1) + 2j \nu_I = 2(\nu-1)$. This agrees with Fermat's principle where the incremental phase change of a ray is directly related to the product of the path length along the ray and refractive index contrast $(\nu(\bm{r}) -1)$. In other words, the incremental phase change of a ray per wavelength should be proportional to $k_0(\nu-1)$. For conventional RA, it is known (using asymptotic techniques) that the incremental phase change per wavelength is $\frac{1}{2} k_0 (\nu^2 -1)$ which does not match the expected phase change as per Fermat's principal (\ref{Eq_rytovfulldB6-1}). Therefore xPRA-LM also appears to better satisfy the underlying physics of the problem.

\section{Distorted Wave Extended Phaseless Rytov Iterative Method (DxPRIM)}
\label{Sec_DxPRIM_LM}

\begin{figure*}[!h]
	\centering
	\includegraphics[width=6.3in]{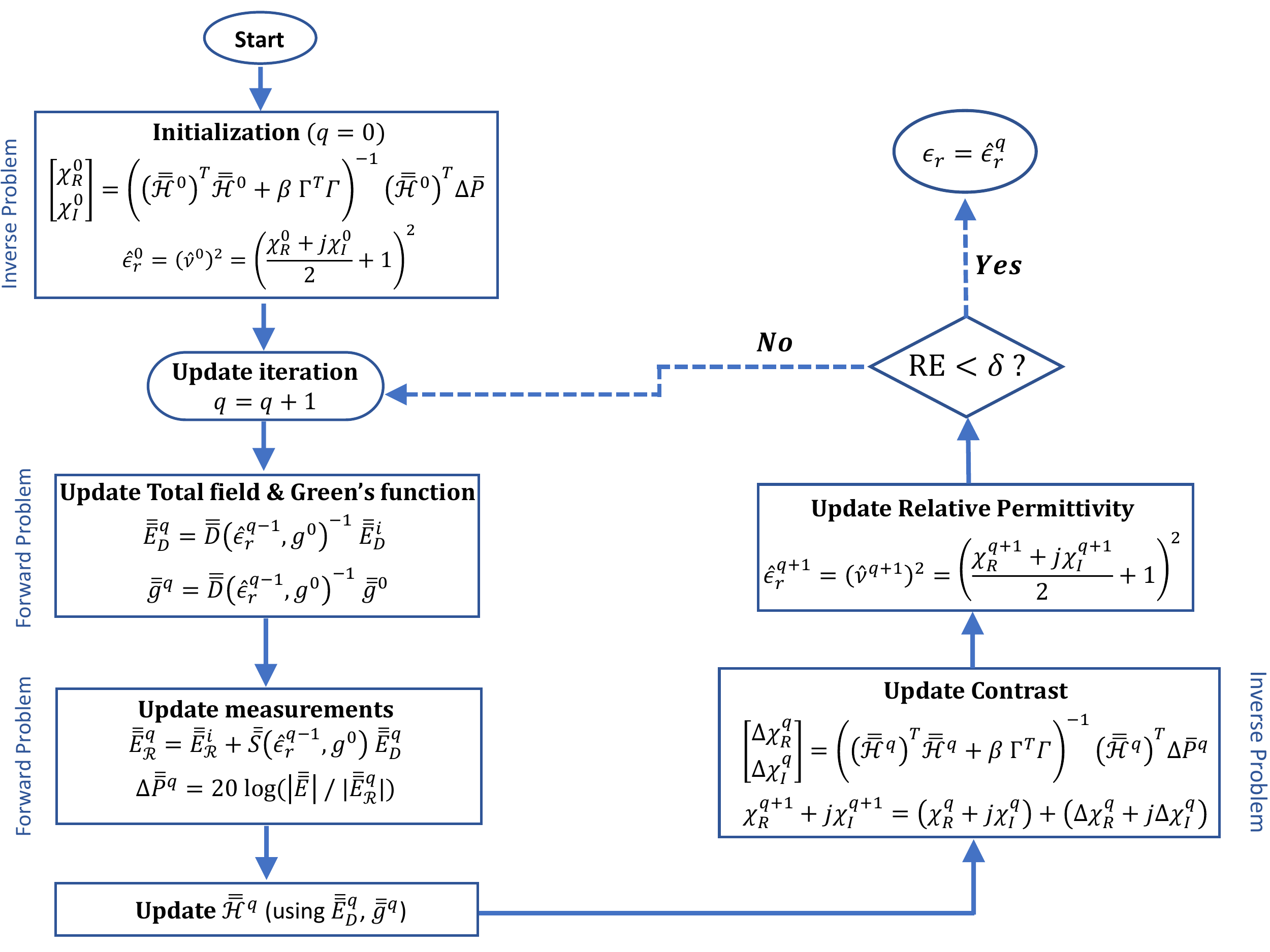}
	\caption{Flow chart for DxPRIM. Its detailed operation is described in the text.}
	\label{flowchart}
	\vspace{-0.2\baselineskip}
\end{figure*}

DxPRIM uses DxPRM (\ref{Eq_DW_xPRA9_reg}) iteratively for updating the contrast profile. The forward problem equations (\ref{Eq_VSI_MV}) and (\ref{Eq_VSI1_MV}), are used for updating the total field (inside $\mathcal{D}$ and at receivers) using an estimate of the contrast profile given by DxPRM (\ref{Eq_DW_xPRA9_reg}) (or (\ref{Eq_DW_xPRA8_reg})). The inhomogeneous Green's function also needs to be updated iteratively. It can be estimated as a solution of the forward state problem ((\ref{Eq_VSI}) or (\ref{Eq_VSI_MV})). Let $g^b$ be the inhomogeneous Green's function for any given background profile with relative permittivity profile $\epsilon_r^b$. We can use Lippmann-Schwinger equation (\ref{Eq_VSI}) to estimate $g^b$ at a point $\bm{r}$ inside $\mathcal{D}$ due to a fictitious source at the receiver location $\bm{r}_{m_r}$ as,
\begin{equation}
	\label{Eq_inhomo_gf}
	\begin{aligned}
	& g^b(\bm{r}_{m_r}, \bm{r})  = \\ & E^i(\bm{r}_{m_r}, \bm{r}) + k_0^2 \int_{\mathcal{D}} g^0(\bm{r}, \bm{r}') (\epsilon_r^b(\bm{r}')-1) g^b(\bm{r}', \bm{r}_{m_r}) d\bm{r}'^2
	\end{aligned}
\end{equation}
where, $E^i(\bm{r}_{m_r}, \bm{r})$ is the free-space Green's function $g^0(\bm{r}_{m_r}, \bm{r})$ such that we are estimating the incident field at point $\bm{r}$ due to a source at $\bm{r}_{m_r}$. Equation (\ref{Eq_inhomo_gf}) can be solved using MoM (similar to solving (\ref{Eq_VSI}) using MoM to obtain (\ref{Eq_VSI_MV})),
\begin{equation}
\label{Eq_VSI_GF}
\rttensortwo{g}^0 = {\rttensortwo{D}}(\epsilon_r^b, g^0) \cdot \rttensortwo{g}^b
\end{equation}
where, $\rttensortwo{{g}}^0 \in \mathbb{C}^{N\times M_r}$ is the free-space homogeneous Green's function and $\rttensortwo{g}^b \in \mathbb{C}^{N\times M_r}$ is the inhomogeneous Green's function for the background with permittivity profile $\epsilon_r^b$. Matrix $\rttensortwo{{D}} \in \mathbb{C}^{N\times N}$ is the forward state matrix (\ref{Eq_VSI_MV}).

Finally, using the three forward problem equations (\ref{Eq_VSI_MV}), (\ref{Eq_VSI1_MV}) and (\ref{Eq_VSI_GF}) and two inverse problem equations (\ref{Eq_DW_xPRA9_reg}) and (\ref{Eq_DW_xPRA8_reg}), we can now formulate the distorted extended phaseless Rytov iterative method (DxPRIM). Fig. \ref{flowchart} provides the flowchart for the DxPRIM algorithm which can be explained step-by-step as follows:
\begin{enumerate}[leftmargin=0.5cm]
	\item \textbf{Initialize Relative Permittivity} (Step $q=0$): The initial guess for contrast profile $[\rttensorone{\chi}^{0}_{{R}}\ \ \rttensorone{\chi}^{0}_{{I}}]^T \in \mathbb{R}^{2N\times 1}$ is obtained from (\ref{Eq_DW_xPRA8_reg}). This is the solution to the modified xPRA-LM (in (\ref{Eq_xPRA})). The initial complex-valued relative permittivity profile $\rttensorone{\hat{\epsilon}}_r^{0} \in \mathbb{C}^{N\times 1}$ is then obtained as,
	\begin{equation}
		\label{Eq_flow1}
		\rttensorone{\hat{\epsilon}}_r^{0} = \left(\frac{\rttensorone{\chi}^{0}_{{R}}+ j \rttensorone{\chi}^{0}_{{I}}}{2} + 1 \right)^2
	\end{equation}

	\item \textbf{Update total field and inhomogeneous Green's function}: The estimate of the permittivity profile $\rttensorone{\hat{\epsilon}}_r^q$ is used as the inhomogeneous background. Using this estimate of permittivity profile, the forward problem (\ref{Eq_VSI_MV}) is solved in order to update the total field $\rttensortwo{E}_{\scaleto{\mathcal{D}}{3pt}}^q$ inside $\mathcal{D}$,
	\begin{equation}
	\label{Eq_VSI_MV_it}
	{\rttensortwo{E}_{\scaleto{\mathcal{D}}{3pt}}^q = {\rttensortwo{D}}(\hat{\epsilon}_r^{q-1}, g^0)}^{-1}  \rttensortwo{E}^i_{\scaleto{\mathcal{D}}{3pt}}
	\end{equation}
	Similarly, the inhomogeneous Green's function is estimated inside $\mathcal{D}$ using the permittivity profile estimate as
	\begin{equation}
	\label{Eq_VSI_GF_it}
	\rttensortwo{g}^b= {{\rttensortwo{D}}(\hat{\epsilon}_r^{q-1}, g^0)}^{-1} \rttensortwo{g}^0
	\end{equation}

	\item \textbf{Update Measurements and Inverse Model Matrix}: Using the estimate of the total field $\rttensortwo{E}^q_{\scaleto{\mathcal{D}}{3pt}}$ inside $\mathcal{D}$, the total field at the receiver is estimated as,
	\begin{equation}
	\label{Eq_VSI1_MV_it}
	\rttensortwo{{E}}^q_{\scaleto{\mathcal{R}}{3pt}} = \rttensortwo{{E}}^i_{\scaleto{\mathcal{R}}{3pt}} + \rttensortwo{{S}}(\hat{\epsilon}_r^{q-1}, g^0) \cdot \rttensortwo{E}^q_{\scaleto{\mathcal{D}}{3pt}}.
	\end{equation}
	In addition, update the change in RSS vector by using the current estimate of permittivity profile ($\hat{\epsilon}_r^q$) as background as,
	\begin{equation}
	\label{Eq_RSS_it}
	\begin{aligned}
		\Delta \rttensortwo{P}^q = 20 \log \left|\frac{\rttensortwo{E}}{{\rttensortwo{E}^i_{\scaleto{\mathcal{R}}{3pt}}}^q} \right| - 20 \log \left|\frac{\rttensortwo{E}^q_{\scaleto{\mathcal{R}}{3pt}}}{{\rttensortwo{E}^i_{\scaleto{\mathcal{R}}{3pt}}}^q} \right|  &= \Delta \rttensortwo{P} - 20 \log \left|\frac{\rttensortwo{E}^q_{\scaleto{\mathcal{R}}{3pt}}}{{\rttensortwo{E}^i_{\scaleto{\mathcal{R}}{3pt}}}^q} \right|   \\ &= 20 \log \left|\frac{\rttensortwo{E}}{\rttensortwo{E}^q_{\scaleto{\mathcal{R}}{3pt}}} \right|
	\end{aligned}
	\end{equation}
	Note that in (\ref{Eq_RSS_it}), $\Delta \rttensortwo{P} = 20 \log \left|\frac{\rttensortwo{E}}{{\rttensortwo{E}^i}^q_{\scaleto{\mathcal{R}}{3pt}}} \right| \in \mathbb{R}^{M_r \times M_t}$ is obtained from measurements and it represents differences in measured total power in the presence of the scatterer (with ground truth profile $\epsilon_r(\bm{r})$) and free-space incident power. Whereas, $\Delta \rttensortwo{P}^q$ represents the difference in measured total power in the presence of the scatterer (with ground truth profile $\epsilon_r(\bm{r})$) and the background total power estimated using $\hat{\epsilon}_r^q(\bm{r})$ as the inhomogeneous background instead of free space. Then, use the total field and the inhomogeneous Green's function estimated in previous step to update the DxPRM model matrix $\rttensortwo{\mathcal{H}}^q$ (as defined in (\ref{Eq_DW_xPRA_H})).
		
	\item \textbf{Update contrast and relative permittivity}: Obtain the change in contrast profile using DxPRM with the total field, model matrix and change in RSS measurements obtained from previous step as,
	\begin{equation}
	\label{Eq_DW_xPRA9_reg_it}
	\begin{aligned}
	\begin{bmatrix}
	{\Delta \rttensorone{\raisedchi}^q_{\tiny R}} \\
	{\Delta \rttensorone{\raisedchi}^q_{\tiny I}} 
	\end{bmatrix} = \left(\big({\rttensortwo{\mathcal{H}}^q}\big)^T \rttensortwo{\mathcal{H}}^q + \beta \Gamma^T \Gamma \right)^{-1} \big({\rttensortwo{\mathcal{H}}^q}\big)^T \Delta \rttensorone{P}^q
	\end{aligned}
	\end{equation}
	where $\Delta \rttensorone{P}^q \in \mathbb{R}^{M\times1}$ is the unrolled form of matrix $\Delta \rttensortwo{P}^q \in \mathbb{R}^{M_r\times M_t}$ which is estimated in the previous step. Use this change in contrast estimated from (\ref{Eq_DW_xPRA9_reg_it}) to update the contrast for next iteration,
	\begin{equation}
		\begin{aligned}
		\rttensorone{\chi}^{q+1}_{\tiny R} + j \rttensorone{\chi}^{q+1}_{\tiny I} = \left(\rttensorone{\chi}^{q}_{\tiny R} + j \rttensorone{\chi}^{q}_{\tiny I}\right) + \left(\Delta \rttensorone{\chi}^{q}_{\tiny R} + j \Delta\rttensorone{\chi}^{q}_{\tiny I}\right)
		\end{aligned}
	\end{equation}
	which can be used to obtain updated relative permittivity profile for next iteration,
	\begin{equation}
		\begin{aligned}
		\rttensorone{\hat{\epsilon}}_r^{q+1} = \left(\rttensorone{\hat{\nu}}^{q+1} \right)^2 = \left(\frac{\rttensorone{\chi}^{q+1}_{\tiny R} + j \rttensorone{\chi}^{q+1}_{\tiny I}}{2} + 1\right)^2
		\end{aligned}
	\end{equation}
	
	\item \textbf{Terminate or continue to next iteration}: Using the current estimate of relative permittivity $\hat{\epsilon}_r^q$, repeat steps 2-4 until the estimate of relative permittivity $\hat{\epsilon}_r^q$ converges.  Since the ground truth permittivity profile  ${\epsilon}_r$ is unknown, the decision of continuing or stopping can be based on checking if the current estimate of the magnitude of the total field $|{\rttensortwo{E}^q_{\scaleto{\mathcal{R}}{3pt}}}|$ at the receivers is close to the actual measured magnitude of total field $| \rttensortwo{E} |$ at the receivers. We define the relative error between the measured and estimated magnitude of total field as,
	\begin{equation}
		\begin{aligned}
		\text{RE}= \left| \frac{\left| \rttensorone{E}\right| - \left|\rttensorone{E}^q_{\scaleto{\mathcal{R}}{3pt}}\right|}{\left| \rttensorone{E}\right|} \right|_1
	\end{aligned}
	\end{equation}
	where $|\cdot|_1$ is a norm 1 operator and $\rttensorone{E}$ and $\rttensorone{E}^q_{\scaleto{\mathcal{R}}{3pt}} \in \mathbb{C}^{M\times 1}$ are unrolled form of matrices $\rttensortwo{E}$ and $\rttensortwo{E}^q_{\scaleto{\mathcal{R}}{3pt}}  \in \mathbb{C}^{M_r \times M_t}$ respectively. We can define a threshold $\delta$ and if RE $< \delta $, the iterative algorithm is terminated. The value of $\delta$ generally depends on the type of the profile. If the profile is smaller than a wavelength, then $\delta$ can be extremely small (we use $\delta = 0.01$), else it can be slightly larger (we use $\delta = 0.1$). 
\end{enumerate}

\section{Simulation and Experimental Results}
\label{Sec_results}
In this section we provide simulation and experimental results to demonstrate the performance of the proposed DxPRIM technique. In the results the frequency utilized is 2 GHz unless otherwise stated. 

For performance benchmarking, we compare DxPRIM to the state-of-the-art phaseless inverse scatting technique \cite{Xudongchen, chen2010subspace, chen2018computational}, subspace optimization (SOM), and it is denoted PD-SOM. Recently, PD-SOM's results have been further improved using deep learning \cite{Xudongchen}, however, we only provide comparisons with the original PD-SOM model (without deep learning) so that comparisons between DxPRIM and PD-SOM can be obtained without learning features from large data-sets. The implementation details for PD-SOM can be found in \cite{chen2018computational, Xudongchen}. 
\begin{figure}[!h]
	\captionsetup[subfigure]{justification=centering}
	\centering
	%%%%%%%%%%%%%%%%%%%%%%%%%%%%%%%
	\begin{subfigure}[t]{0.21\textwidth}
		\centering
		\includegraphics[width=\textwidth]{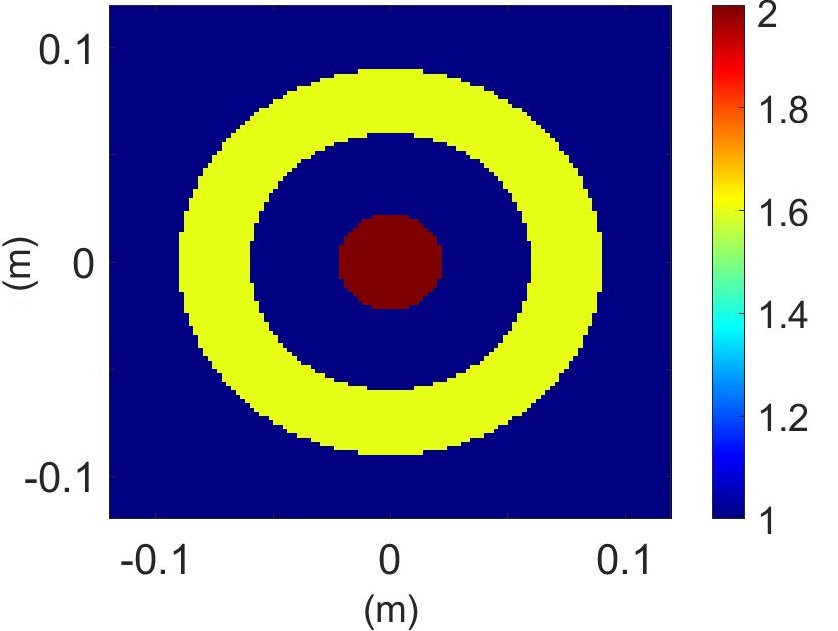}
		\subcaption*{Ground Truth: $\operatorname*{Re}(\epsilon_r)$}
	\end{subfigure}       \hspace{1mm}
	\begin{subfigure}[t]{0.21\textwidth}
		\centering
		\includegraphics[width=\textwidth]{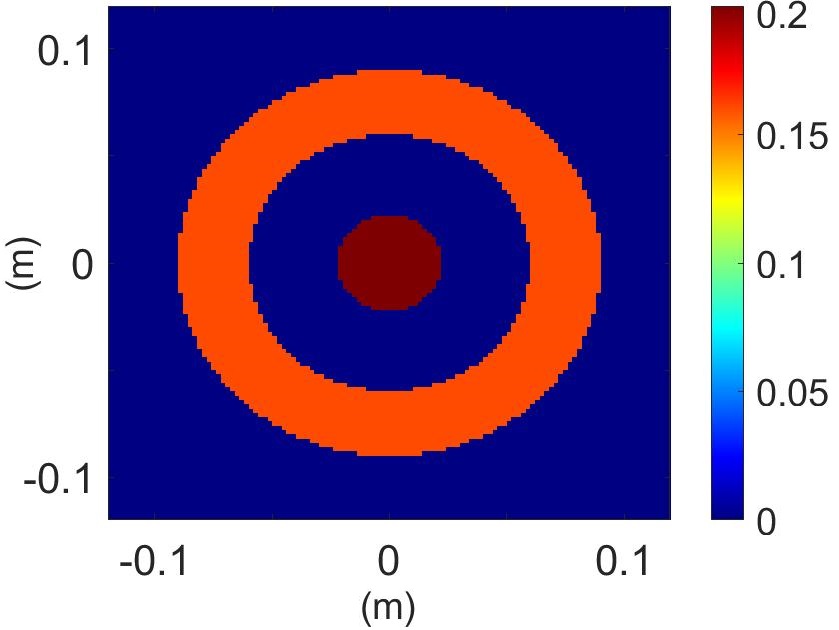}
		\subcaption*{Ground Truth: $\operatorname*{Im}(\epsilon_r)$}
	\end{subfigure} 
	%%%%%%%%%%%%%%%%%%%%%%%%%%%%%%%%%%%
	\caption{Ground truth scatterer profile with a circular disc enclosed inside a ring.  A circular cylinder with $\epsilon_r = 2+0.2j$ is placed inside a concentric annulus with $\epsilon_r = 1.6+0.16j$.}
	\label{GTringcircle} 
	\vspace{-0.5\baselineskip}	
\end{figure}

\subsection{Simulation Results}

\begin{figure}[!h]
	\captionsetup[subfigure]{justification=centering}
	\centering
	%%%%%%%%%%%%%%%%%%%%%%%%%%%%%%%
	\adjustbox{minipage=1em,raise=6ex}{\subcaption{ }\label{}}% 
	\begin{subfigure}[t]{0.2\textwidth}
		\centering
		\includegraphics[width=1.3in]{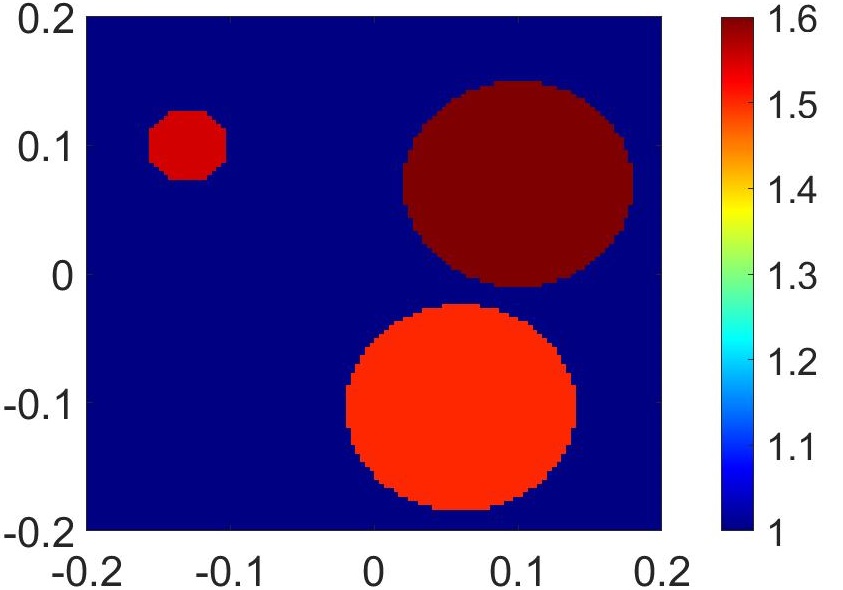}
		\subcaption*{Ground Truth: $\operatorname*{Re}(\epsilon_r)$}
	\end{subfigure}       \hspace{3mm}%
	\begin{subfigure}[t]{0.2\textwidth}
		\centering
		\includegraphics[width=1.3in]{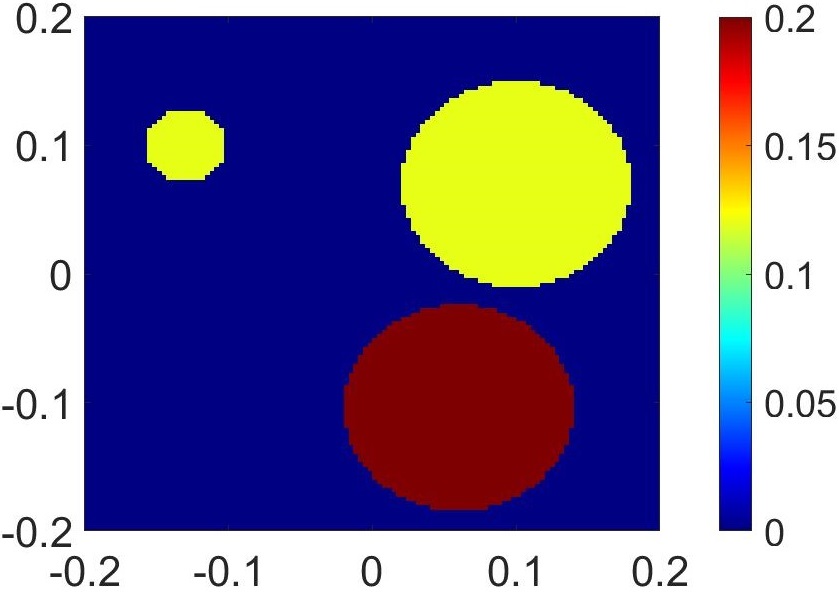}
		\subcaption*{Ground Truth: $\operatorname*{Im}(\epsilon_r)$}
	\end{subfigure} 
	\par\smallskip  
	\adjustbox{minipage=1em,raise=6ex}{\subcaption{ }\label{}}% 
	\begin{subfigure}[t]{0.2\textwidth}
		\centering
		\includegraphics[width=1.3in]{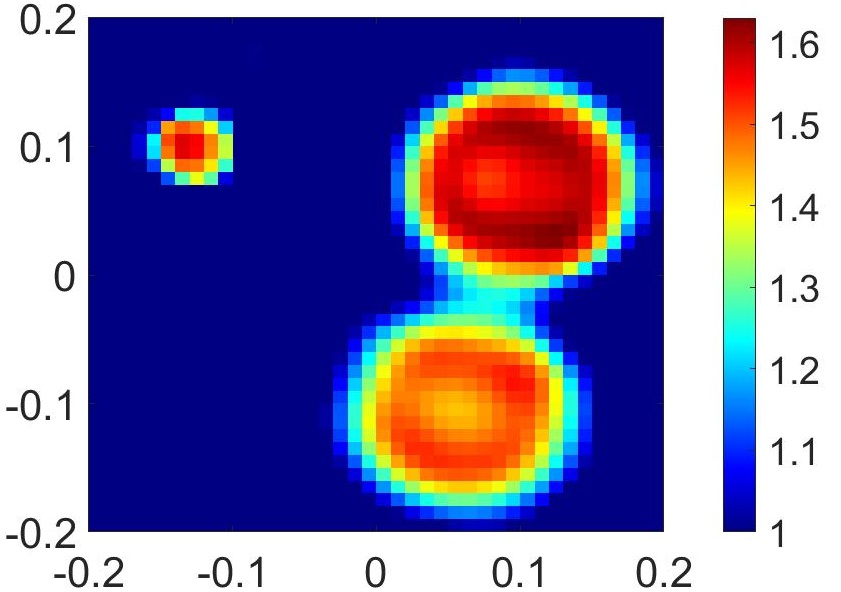}
		\subcaption*{DxPRIM $\operatorname*{Im}(\epsilon_r)$ \\(PSNR=32 dB, SSIM=0.94)}
	\end{subfigure}       \hspace{3mm}%
	\begin{subfigure}[t]{0.2\textwidth}
		\centering
		\includegraphics[width=1.3in]{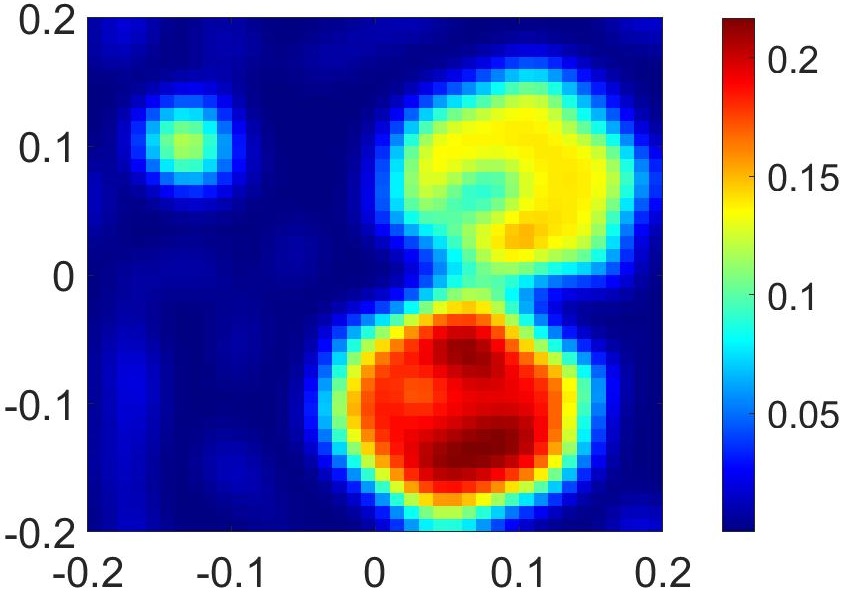}
		\subcaption*{DxPRIM $\operatorname*{Im}(\epsilon_r)$ \\(PSNR=24 dB, SSIM=0.85)}
	\end{subfigure}
	\par\smallskip
	\adjustbox{minipage=1em,raise=6ex}{\subcaption{ }\label{}}%   
	\begin{subfigure}[t]{0.2\textwidth}
		\centering
		\includegraphics[width=1.3in]{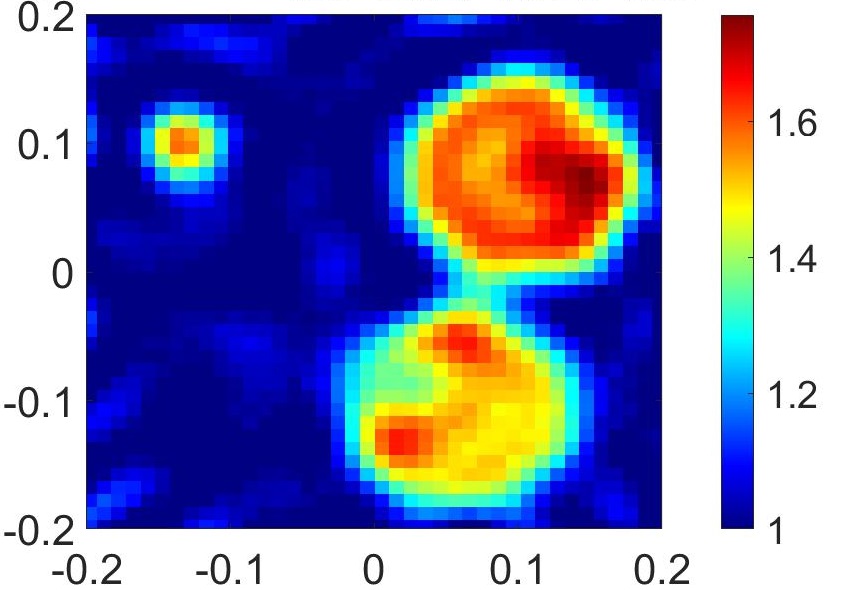}
		\subcaption*{PD-SOM $\operatorname*{Im}(\epsilon_r)$ \\(PSNR=22 dB, SSIM=0.88)}
	\end{subfigure}       \hspace{3mm}%
	\begin{subfigure}[t]{0.2\textwidth}
		\centering
		\includegraphics[width=1.3in]{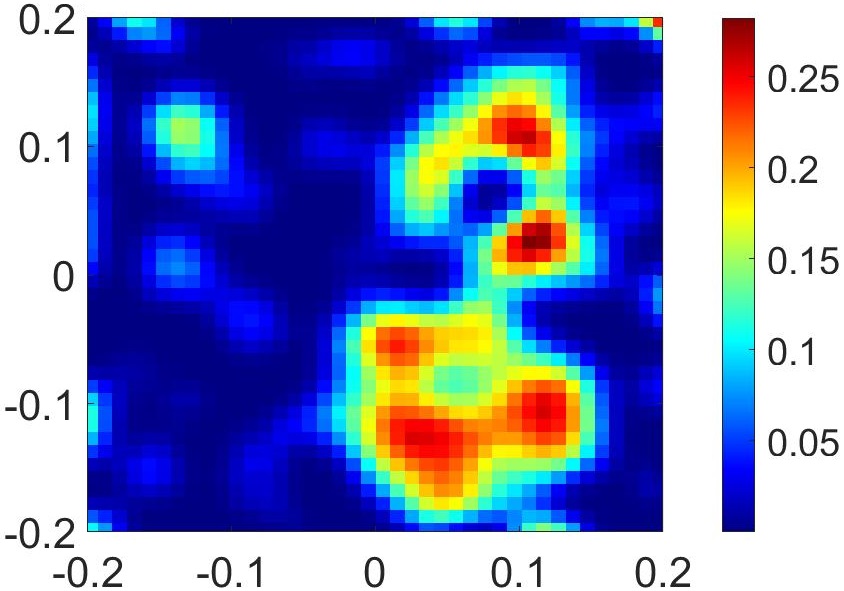}
		\subcaption*{PD-SOM $\operatorname*{Im}(\epsilon_r)$ \\(PSNR=14 dB, SSIM=0.80)}
	\end{subfigure} 
	\par\smallskip  
	%%%%%%%%%%%%%%%%%%%%%%%%%%%%%%%%%%%
	\caption{Reconstructions for an eccentric cylinder structure. (a) Ground truth scatterer profile. The scatterers on top-left, top-right and bottom-right respectively have $\epsilon_r=1.55+0.12j, 1.6+0.12j$ and $1.5+0.2j$. (b) reconstruction using DxPRIM, (c) reconstruction using PD-SOM. x-axis and y-axis are in meters.}
	\label{2021_deep_learn} 
	\vspace{-0.5\baselineskip}	
\end{figure}

\begin{figure}[!h]
	\captionsetup[subfigure]{justification=centering}
	\centering
	%%%%%%%%%%%%%%%%%%%%%%%%%%%%%%%
	\adjustbox{minipage=1em,raise=6ex}{\subcaption{ }\label{}}% 
	\begin{subfigure}[t]{0.2\textwidth}
		\centering
		\includegraphics[width=1.3in]{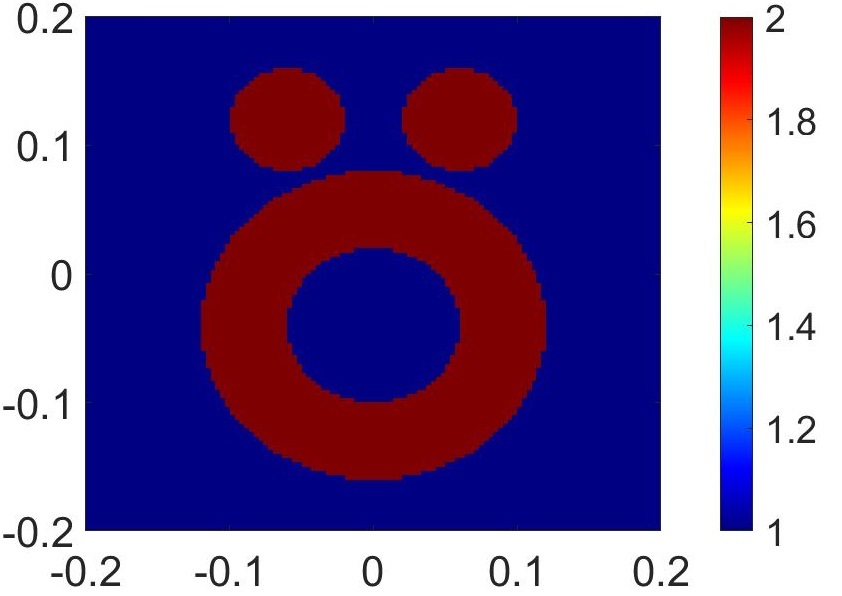}
		\subcaption*{Ground Truth: $\operatorname*{Re}(\epsilon_r)$}
	\end{subfigure}       \hspace{2mm}%
	\begin{subfigure}[t]{0.2\textwidth}
		\centering
		\includegraphics[width=1.3in]{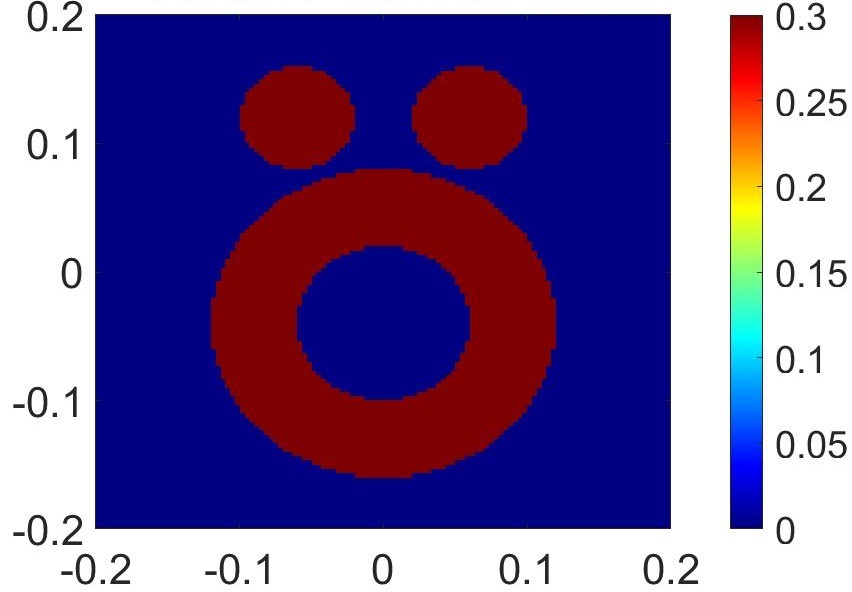}
		\subcaption*{Ground Truth: $\operatorname*{Im}(\epsilon_r)$}
	\end{subfigure} 
	\par\smallskip  
	\adjustbox{minipage=1em,raise=6ex}{\subcaption{ }\label{}}% 
	\begin{subfigure}[t]{0.2\textwidth}
		\centering
		\includegraphics[width=1.3in]{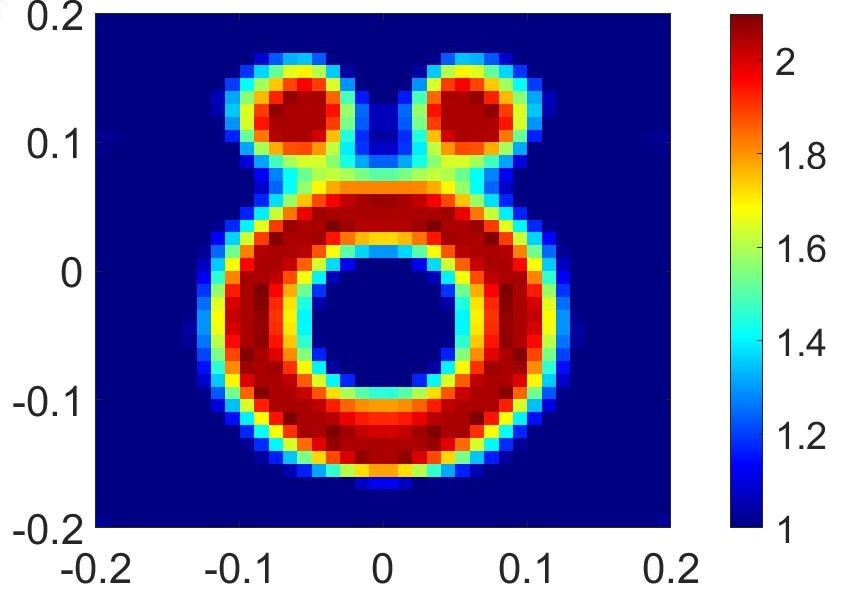}
		\subcaption*{DxPRIM $\operatorname*{Re}(\epsilon_r)$ \\(PSNR=24 dB, SSIM=0.97)}
	\end{subfigure}       \hspace{2mm}%
	\begin{subfigure}[t]{0.2\textwidth}
		\centering
		\includegraphics[width=1.3in]{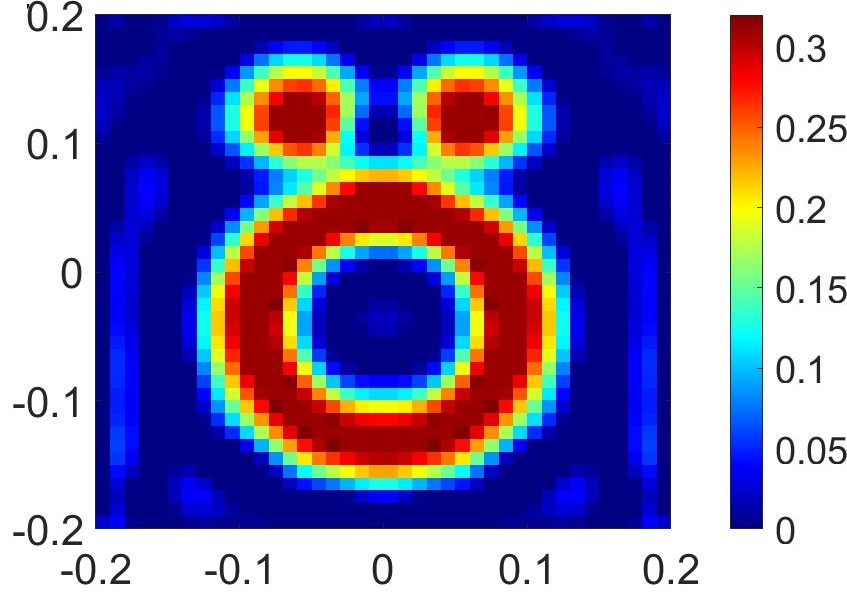}
		\subcaption*{DxPRIM $\operatorname*{Im}(\epsilon_r)$ \\(PSNR=20 dB, SSIM=0.94)}
	\end{subfigure} 
	\par\smallskip  
	\adjustbox{minipage=1em,raise=6ex}{\subcaption{ }\label{}}% 
	\begin{subfigure}[t]{0.2\textwidth}
		\centering
		\includegraphics[width=1.3in]{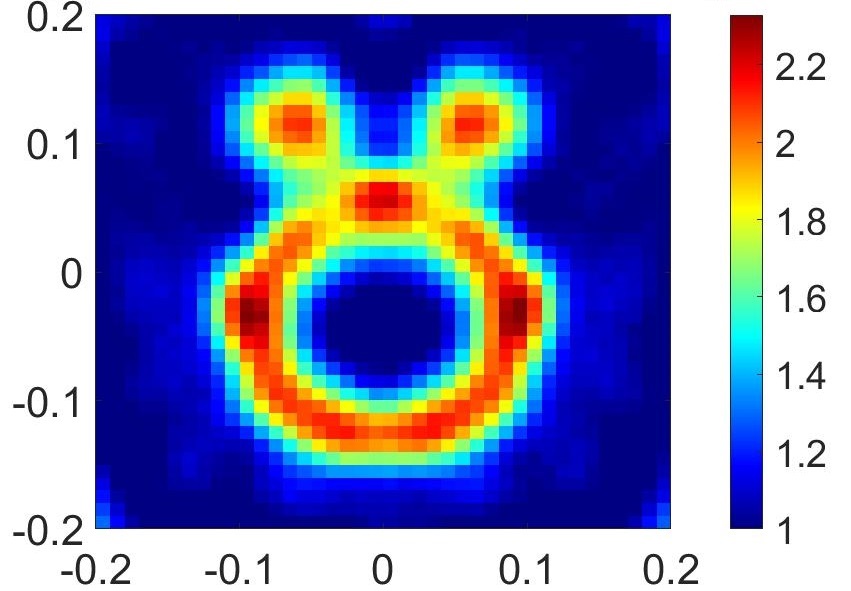}
		\subcaption*{PD-SOM $\operatorname*{Re}(\epsilon_r)$ \\(PSNR=15 dB, SSIM=0.84)}
	\end{subfigure}       \hspace{2mm}%
	\begin{subfigure}[t]{0.2\textwidth}
		\centering
		\includegraphics[width=1.3in]{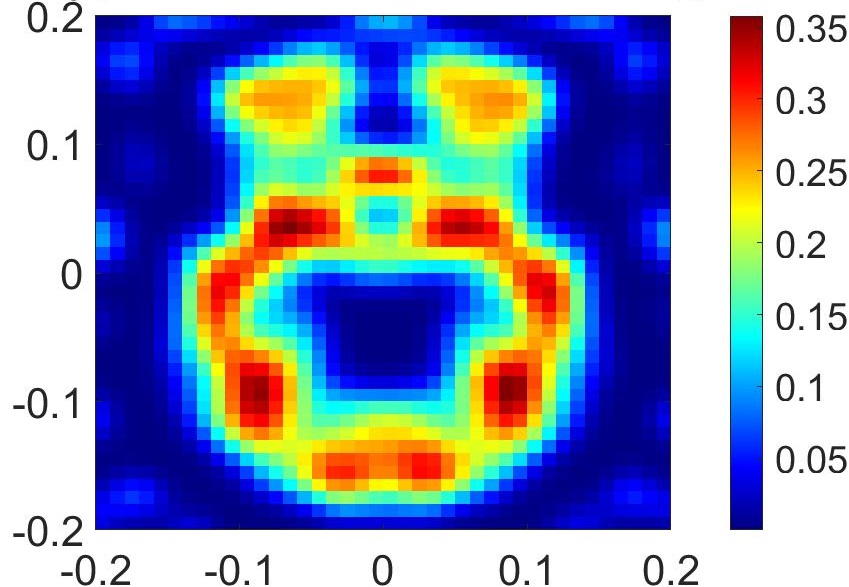}
		\subcaption*{PD-SOM $\operatorname*{Im}(\epsilon_r)$ \\(PSNR=9 dB, SSIM=0.78)}
	\end{subfigure} 
	\par\smallskip  
	%%%%%%%%%%%%%%%%%%%%%%%%%%%%%%%%%%%
	\caption{``Austria" profile reconstructions. (a) Ground truth ``Austria" profile with $\epsilon_r=2+0.3j$, (b) Reconstruction using DxPRIM, (c) Reconstruction using PD-SOM. x-axis and y-axis are in meters.}
	\label{Austria} 
	\vspace{-0.5\baselineskip}	
\end{figure}

We consider three scatterer profiles for reconstruction. These are profiles from other state-of-the-art PD-ISP work, including PD-SOM and its deep learning extension \cite{Xudongchen, chen2010subspace, chen2020review, chen2018computational}. These three profiles are defined as:
\begin{itemize}
	\item A concentric ring structure as used in PD-SOM \cite{chen2010subspace, chen2018computational} and it is shown in Fig. \ref{GTringcircle}. A circular cylinder with $\epsilon_r = 2+0.2j$ is placed inside a concentric annulus with $\epsilon_r = 1.6+0.16j$.
	\item Three eccentric circular scatterers with different complex-valued permittivity values and size as  shown in Fig. \ref{2021_deep_learn}(a). It is an electrically large, homogeneous lossy scatterer profile utilized previously \cite{Xudongchen}.
	\item The well known ``Austria" profile as shown in Fig. \ref{Austria}(a) \cite{Xudongchen, chen2010subspace, chen2018computational, chen2020review, TGRS2, 8709721}.
\end{itemize}

To quantify the performance of the techniques, we use peak signal-to-noise ratio (PSNR) and Structural Similarity Index (SSIM) \cite{ssim2004} metrics to compare the ground truth image to the reconstructed image. The higher the PSNR value, the better is the reconstruction. SSIM value ranges from 0 to 1 and the closer to unity, the better is the reconstruction. %Typically PSNR values greater than 10-30 dB and SSIM values greater than 0.85 are considered very good.

For all the simulation results we utilize a uniform measurement setup consisting of $40$ transceivers placed around $\mathcal{D}$ on a circular boundary of radius $5 \lambda_0$. When one transceiver acts as a transmitter, all other 39 transceiver act as receivers. Hence there are in total $M = 40(40-1)/2 = 780$ unique measurements for the RSS values. For reconstruction, $\mathcal{D}$ is divided into $40 \times 40$ discrete grids. Hence we need to estimate $2N = 3200$ unknowns using $780$ measurements. To generate measurement data (magnitude of total field, $|E|$), the forward problem is solved using MoM (as outlined in section \ref{Sec_prob_form}). For the forward problem, $\mathcal{D}$ is divided into $120 \times 120$ grids. We also add 5\% AWGN noise $\overline{n}$ to the simulated magnitude of the total field ($|\overline{E}|+\overline{n}$) where $|\overline{E}| \in \mathbb{R}^{M\times 1}$ and the level of noise is defined as $||\overline{n}||/||\overline{E}||$, where $||\cdot||$ is the Euclidean norm.

\begin{figure*}
	\captionsetup[subfigure]{justification=centering}
	\centering
	\adjustbox{minipage=3em}{\subcaption*{ }}
	\adjustbox{minipage=0.2\textwidth}{\subcaption*{\textbf{{\small DxPRIM $\operatorname*{Re}(\epsilon_r)$}} }}%
	\adjustbox{minipage=0.2\textwidth}{\subcaption*{\textbf{{\small DxPRIM $\operatorname*{Im}(\epsilon_r)$}} }}\hspace{12mm}% 
	\adjustbox{minipage=0.2\textwidth}{\subcaption*{  \textbf{{\small PD-SOM $\operatorname*{Re}(\epsilon_r)$}}}}%
	\adjustbox{minipage=0.2\textwidth}{\subcaption*{   \textbf{  {\small PD-SOM $\operatorname*{Im}(\epsilon_r)$}}}}\\
	\vspace{-0.2\baselineskip}
	%%%%%%%%%%%%%%%%%%%%%%%%%%%%		
	\adjustbox{minipage=3em,raise=8ex}{\subcaption{\\ 2 GHz}\label{resultsringcirclea}} \hspace{3mm}% 
	\begin{subfigure}[t]{0.2\textwidth}
		\centering
		\includegraphics[scale=0.17]{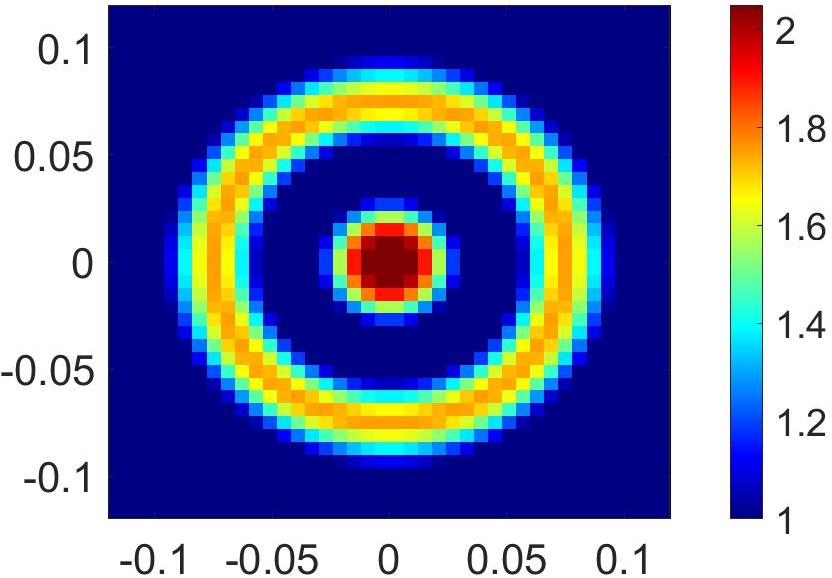}
		\caption*{(PSNR=30.5 dB, SSIM=0.91)}
	\end{subfigure}\hspace{2mm}%
	\begin{subfigure}[t]{0.2\textwidth}
		\centering
		\includegraphics[scale=0.17]{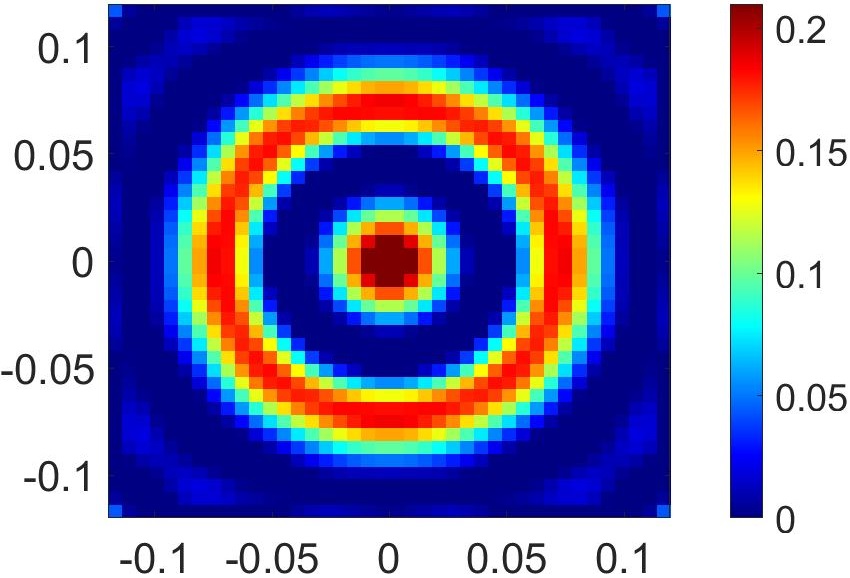}
		\caption*{(PSNR=23 dB, SSIM=0.93)}
	\end{subfigure}\hspace{10mm}% 
	\begin{subfigure}[t]{0.2\textwidth}
		\centering
		\includegraphics[scale=0.17]{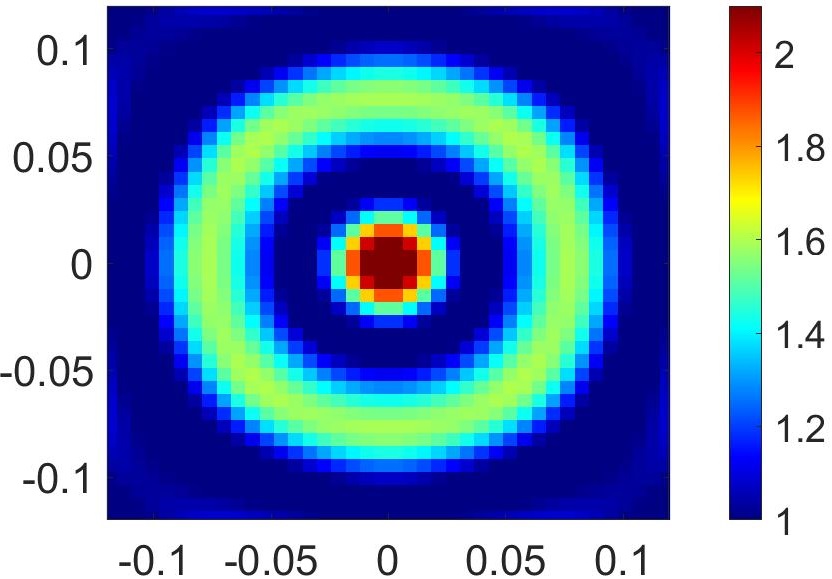}
		\caption*{(PSNR=22 dB, SSIM=0.87)}
		%\caption{bb}
	\end{subfigure}	\hspace{2mm}%
	\begin{subfigure}[t]{0.2\textwidth}
		\centering
		\includegraphics[scale=0.17]{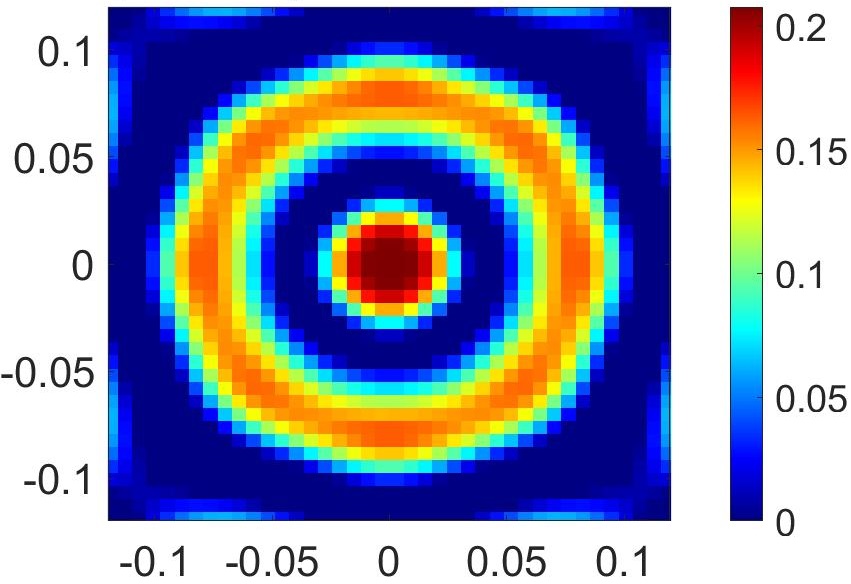}
		\caption*{(PSNR=20 dB, SSIM=0.89)}
		%\caption{bb}
	\end{subfigure}	\\
	\par\smallskip
	%%%%%%%%%%%%%%%%%%%%%%%%%%%%	%%%%%%%%%%%%%%%%%%%%%%%%%%%%				
	\adjustbox{minipage=3em,raise=8ex}{\subcaption{ \\ 4 GHz }\label{resultsringcircleb}} \hspace{3mm}% 
	\begin{subfigure}[t]{0.2\textwidth}
		\centering
		\includegraphics[scale=0.17]{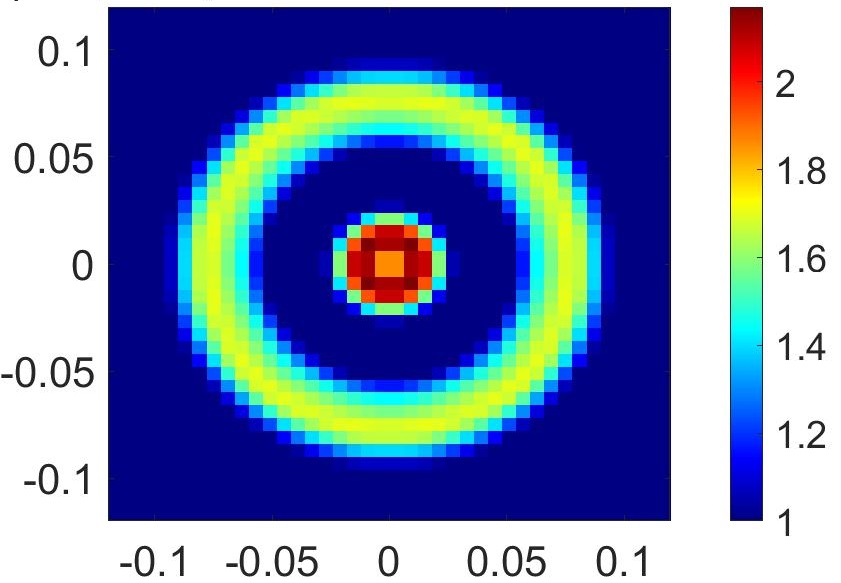}
		\caption*{(PSNR=27 dB, SSIM=0.92)}
	\end{subfigure}\hspace{2mm}%
	\begin{subfigure}[t]{0.2\textwidth}
		\centering
		\includegraphics[scale=0.17]{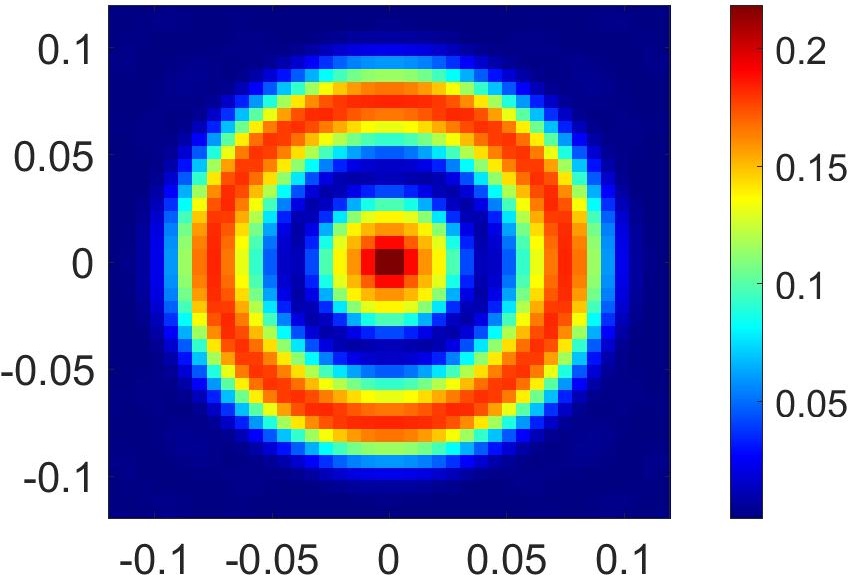}
		\caption*{(PSNR=20 dB, SSIM=0.88)}
	\end{subfigure}\hspace{10mm}% 
	\begin{subfigure}[t]{0.2\textwidth}
		\centering
		\includegraphics[scale=0.17]{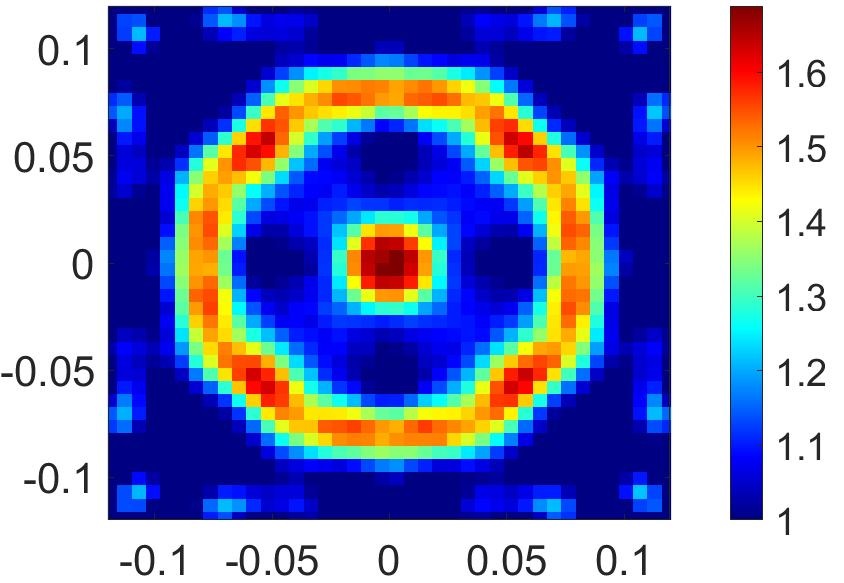}
		\caption*{(PSNR=15 dB, SSIM=0.82)}
		%\caption{bb}
	\end{subfigure}	\hspace{2mm}%
	\begin{subfigure}[t]{0.2\textwidth}
		\centering
		\includegraphics[scale=0.17]{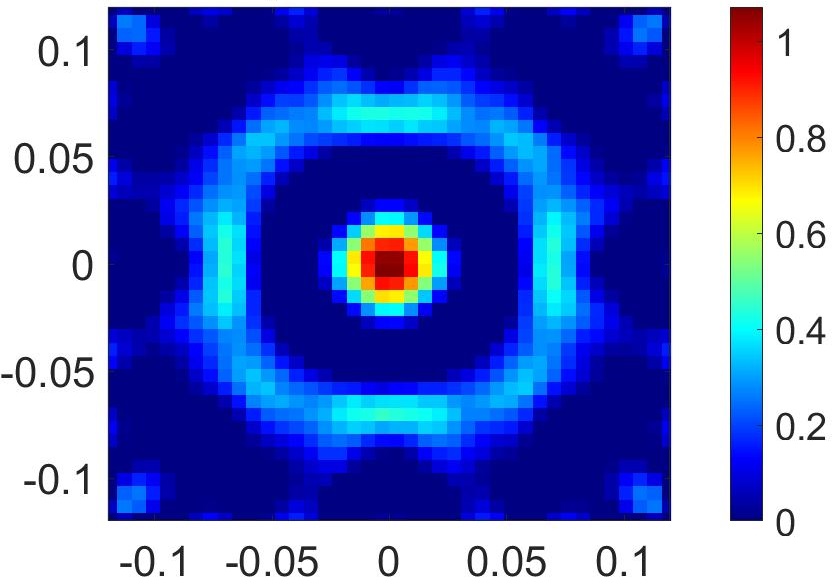}
		\caption*{(PSNR=9 dB, SSIM=0.71)}
		%\caption{bb}
	\end{subfigure}	\\
	\par\smallskip
	%%%%%%%%%%%%%%%%%%%%%%%%%%%%%%%%%%%%%%%%%%%%%%%%%%%%%%%%%%%%%%%%%%%%%%%%%%%%%%%%%%%%		
	\adjustbox{minipage=3em,raise=8ex}{\subcaption{\\ 6 GHz }\label{resultsringcirclec}}\hspace{3mm}% 
	\begin{subfigure}[t]{0.2\textwidth}
		\centering
		\includegraphics[scale=0.17]{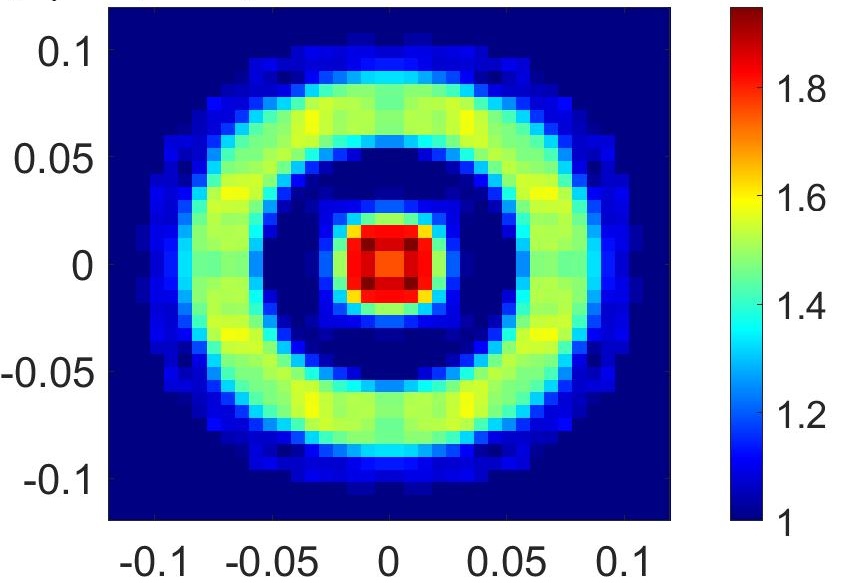}
		\caption*{(PSNR=22 dB, SSIM=0.88)}
	\end{subfigure}\hspace{2mm}%
	\begin{subfigure}[t]{0.2\textwidth}
		\centering
		\includegraphics[scale=0.17]{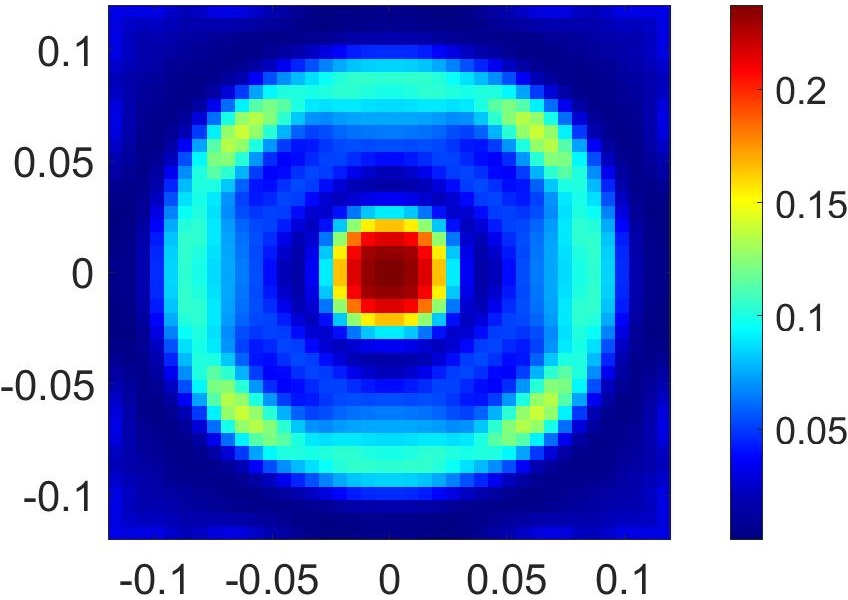}
		\caption*{(PSNR=14 dB, SSIM=0.80)}
	\end{subfigure}\hspace{10mm}% 
	\begin{subfigure}[t]{0.2\textwidth}
		\centering
		\includegraphics[scale=0.17]{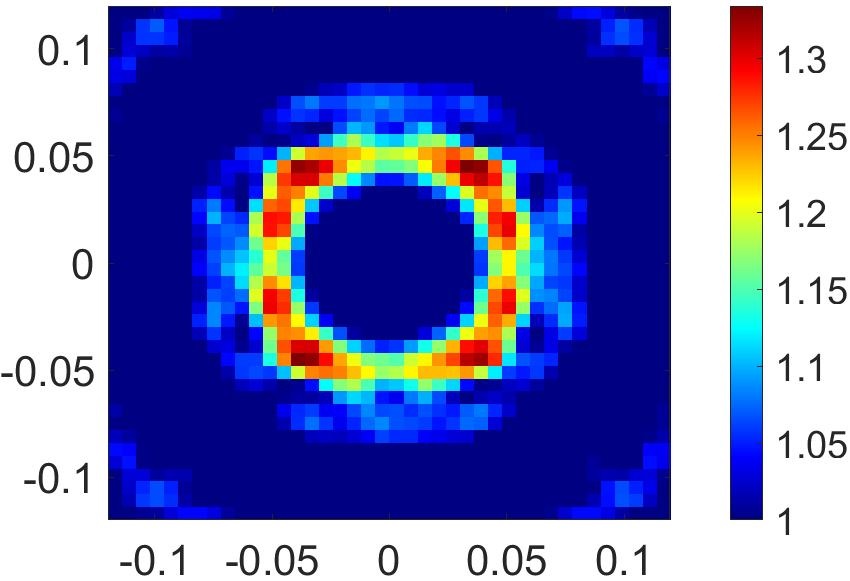}
		\caption*{(PSNR=2 dB, SSIM=0.57)}
		%\caption{bb}
	\end{subfigure}	\hspace{2mm}%
	\begin{subfigure}[t]{0.2\textwidth}
		\centering
		\includegraphics[scale=0.17]{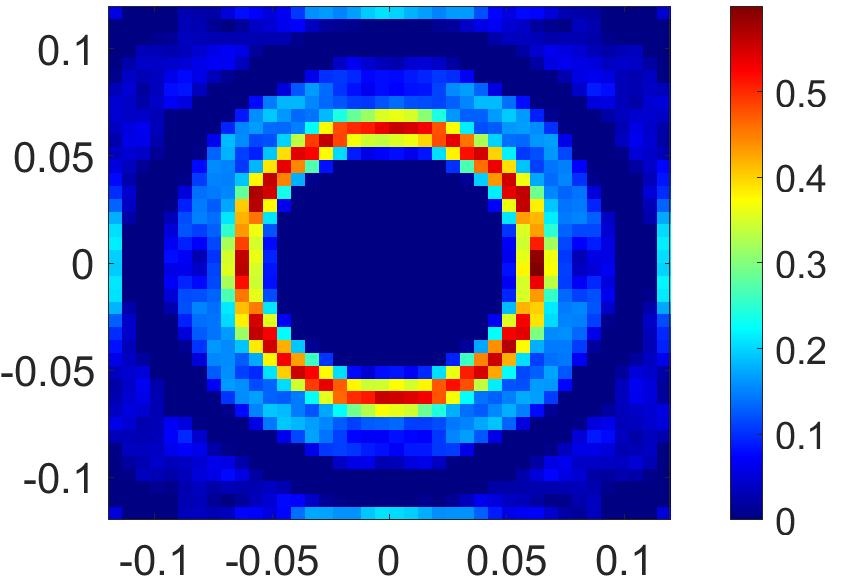}
		\caption*{(PSNR=5 dB, SSIM=0.65)}
		%\caption{bb}
	\end{subfigure}	\\
	\par\smallskip
	%%%%%%%%%%%%%%%%%%%%%%%%%%%%		
	\caption{Reconstruction of the scatterer profile shown in Fig. \ref{GTringcircle} using DxPRIM and PD-SOM. (a), (b) and (c) respectively show reconstruction using 2 GHz, 4 GHz and 6 GHz. First two columns show reconstruction of $\operatorname{Re}(\epsilon_r)$ and $\operatorname{Im}(\epsilon_r)$ profiles respectively using DxPRIM. Last two columns show reconstruction of $\operatorname{Re}(\epsilon_r)$ and $\operatorname{Im}(\epsilon_r)$ profiles respectively using PD-SOM. x-axis and y-axis are in meters.}
	\label{resultsringcircle} 
	\vspace{-0.2\baselineskip}
\end{figure*}

The first scatterer considered is shown in Fig. \ref{GTringcircle} and has been used previously for PD-SOM  \cite{Xudongchen, chen2010subspace, chen2018computational}. The original profile was lossless. Since our method can also handle lossy dielectrics, we also add a loss component to the profile to make it more intricate (keeping the real part of permittivity same). As Fig. \ref{GTringcircle} shows, the scatterer we consider has $\epsilon_r = 2+0.2j$ and is placed inside a concentric annulus with $\epsilon_r = 1.6+0.16j$. We reconstruct this scatterer profile with three frequencies, 2 GHz, 4 GHz, and 6 GHz (corresponding free-space wavelengths respectively are $\lambda_0=0.15$ m, $\lambda_0=0.075$ m, $\lambda_0= 0.05$ m). As frequency increases, the scattering strength of the scatterer increases. %The size of the DoI and scatterer is defined in terms of maximum incident wavelength i.e., $\lambda_0 = 0.15$ m and hence, the size of the DoI is $1.6 \lambda_0 \times 1.6 \lambda_0$ m$^2$. The radius of circular scatterer is $0.15 \lambda_{0}$ and inner and outer radii of concentric annulus is $0.4 \lambda_{0}$ and $0.6 \lambda_{0}$ respectively. 

\begin{figure}[!h]
	\centering
	\includegraphics[width=1.9in]{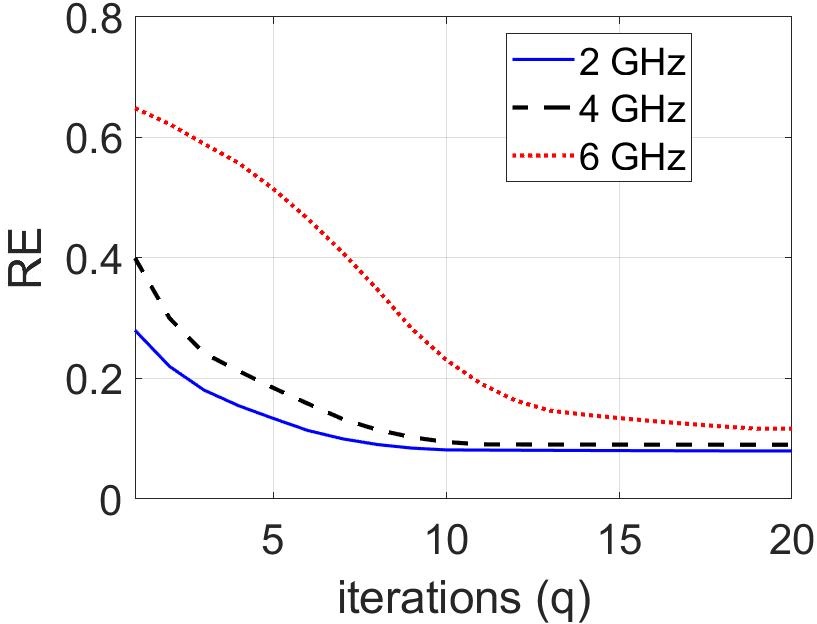}
	\caption{Variation of Relative error (RE) as number of iterations increases in DxPRIM. The plots are for 2 GHz, 4 GHz and 6 GHz reconstructions shown in Fig. \ref{resultsringcircle}. Here RE is relative error between the estimated permittivity and ground truth permittivity.}
	\label{converge}
	\vspace{-0.2\baselineskip}
\end{figure}

Fig. \ref{resultsringcircle} provides reconstruction results for the profile shown in Fig. \ref{GTringcircle} using DxPRIM and PD-SOM. Fig. \ref{resultsringcircle}(a), (b) and (c) respectively provide reconstructions using 2 GHz, 4 GHz and 6 GHz respectively. The first two columns in Fig. \ref{resultsringcircle} provide reconstructions of the real and imaginary part of relative permittivity using the proposed DxPRIM technique. The last two columns provide reconstructions of the real and imaginary parts of relative permittivity using PD-SOM. It can be seen that for 2 GHz, both DxPRIM and PD-SOM can reconstruct the scaterrer. More careful inspection shows that DxPRIM provides better reconstruction and estimation of permittivity compared to PD-SOM (which is reflected in the difference in PSNR and SSIM values shown at the bottom of each reconstruction). Also, DxPRIM converges in just 10 iterations. This can be seen in Fig. \ref{converge} (in solid blue line), which provides the variation of relative error between the estimated and ground truth permittivity profile as the number of iterations in DxPRIM increases.

As frequency is increased to 4 GHz, the difference in performance increases between DxPRIM and PD-SOM.  Specifically, DxPRIM provides significantly better reconstruction of $\operatorname*{Im}(\epsilon_r)$. The convergence is achieved in 12 iterations as can be seen in Fig. \ref{converge} (in dashed black line). Finally, as frequency is further increased to 6 GHz, PD-SOM fails to provide reconstruction whereas DxPRIM still provides good results. As can be seen in Fig. \ref{converge}, DxPRIM requires around 20 iterations to converge which is higher compared to 2 GHz and 4 GHz case, due to stronger scattering at the higher frequency.

Next we select an electrically large, homogeneous lossy scatterer profile used in recent deep learning based PD-SOM work \cite{Xudongchen}. This profile consists of three circular scatterers with different permittivity values and size as shown in Fig. \ref{2021_deep_learn}(a). Similar to previous example, we added 5\% AWGN noise to the measurements. The reconstructions using DxPRIM and PD-SOM at 2 GHz are shown in Fig. \ref{2021_deep_learn}(b) and (c) respectively. Similar to previous results, it can be seen that DxPRIM outperforms PD-SOM, especially in terms of reconstructing $\operatorname*{Im}(\epsilon_r)$ profile.

Finally we provide results for the ``Austria" profile as shown in Fig. \ref{Austria} (a). ``Austria" is a benchmark scatterer profile used in the inverse scattering community \cite{Xudongchen, chen2010subspace, chen2018computational, chen2020review, TGRS2, 8709721} and is also used in state-of-the-art PD-ISP work \cite{Xudongchen, chen2018computational}. It is well-known that the “Austria” profile is a challenging profile for reconstruction \cite{TGRS2}. The physical dimensions of the ``Austria" profile in Fig. \ref{Austria}(a) are defined with respect to 2 GHz ($\lambda_0=0.15$ m). It consists of two disks and one ring. Both of the disks have
a radius of $0.27 \lambda_0$ m and the ring has an inner radius of $0.4 \lambda_0$ m and an exterior
radius of $0.8 \lambda_0$ m. We also add a loss component to it so that the complex-valued relative permittivity becomes $\epsilon_r=2+0.3j$. Similar to previous examples, we added 5\% AWGN noise to the measurements. 

Fig. \ref{Austria}(b) and Fig. \ref{Austria}(c) show reconstruction results using DxPRIM and PD-SOM respectively. It can be clearly seen that DxPRIM outperforms PD-SOM by a large margin, especially in terms of reconstructing $\operatorname*{Im}(\epsilon_r)$ profile. 

\begin{figure}[!h]
	\captionsetup[subfigure]{justification=centering}
	\centering
	%%%%%%%%%%%%%%%%%%%%%%%%%%%%%%%
	\adjustbox{minipage=1em,raise=6ex}{\subcaption{ }\label{}}% 
	\begin{subfigure}[t]{0.2\textwidth}
		\centering
		\includegraphics[width=1.2in]{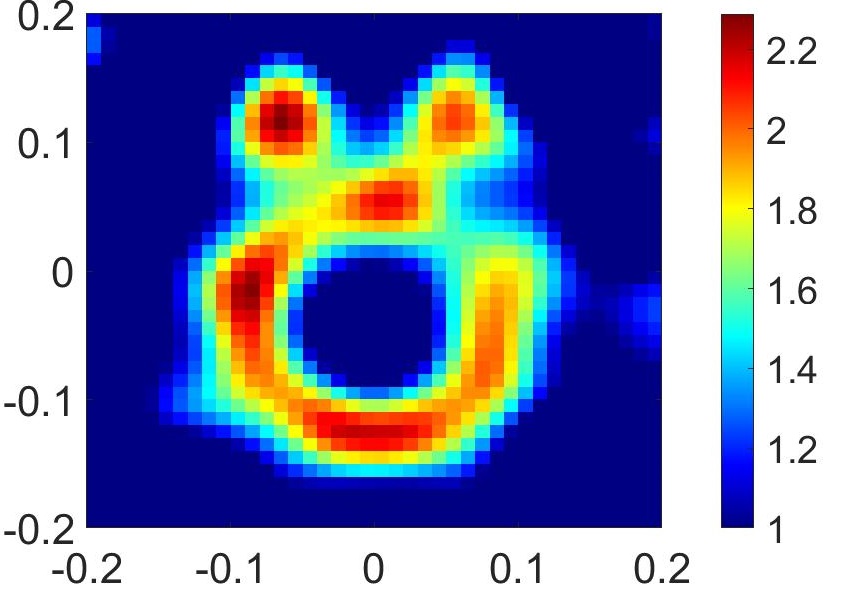}
		\subcaption*{DxPRIM $\operatorname*{Re}(\epsilon_r)$ \\(PSNR=14 dB, SSIM=0.82)}
	\end{subfigure}       %\hspace{3mm}
	\begin{subfigure}[t]{0.2\textwidth}
		\centering
		\includegraphics[width=1.2in]{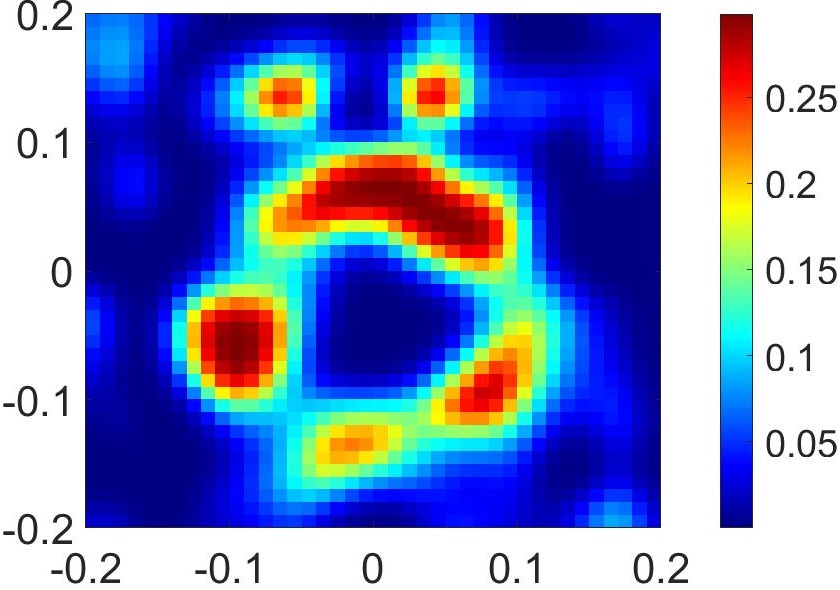}
		\subcaption*{DxPRIM $\operatorname*{Im}(\epsilon_r)$ \\(PSNR=10 dB, SSIM=0.72)}
	\end{subfigure} 
	\par\smallskip  
	\adjustbox{minipage=1em,raise=6ex}{\subcaption{ }\label{}}% 
	\begin{subfigure}[t]{0.2\textwidth}
		\centering
		\includegraphics[width=1.2in]{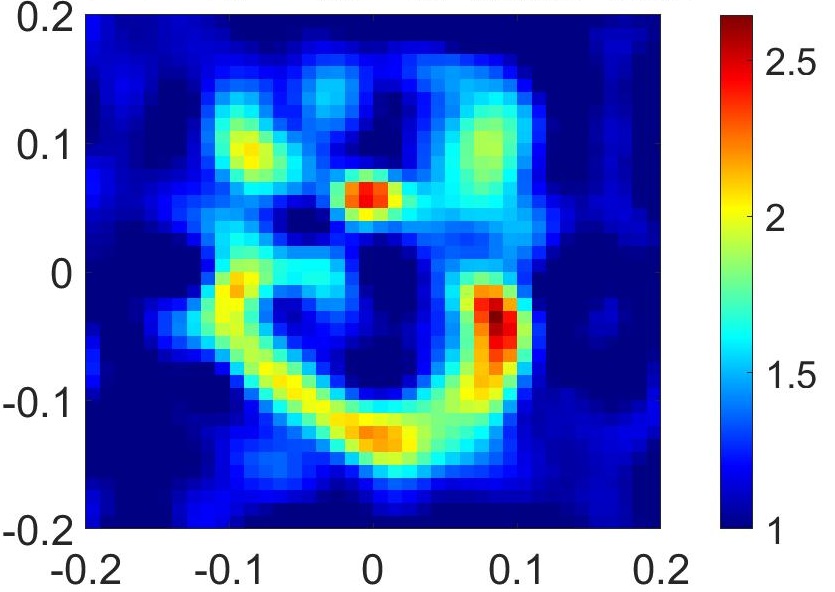}
		\subcaption*{PD-SOM $\operatorname*{Re}(\epsilon_r)$ \\(PSNR=6 dB, SSIM=0.68)}
	\end{subfigure}
	\begin{subfigure}[t]{0.2\textwidth}
		\centering
		\includegraphics[width=1.2in]{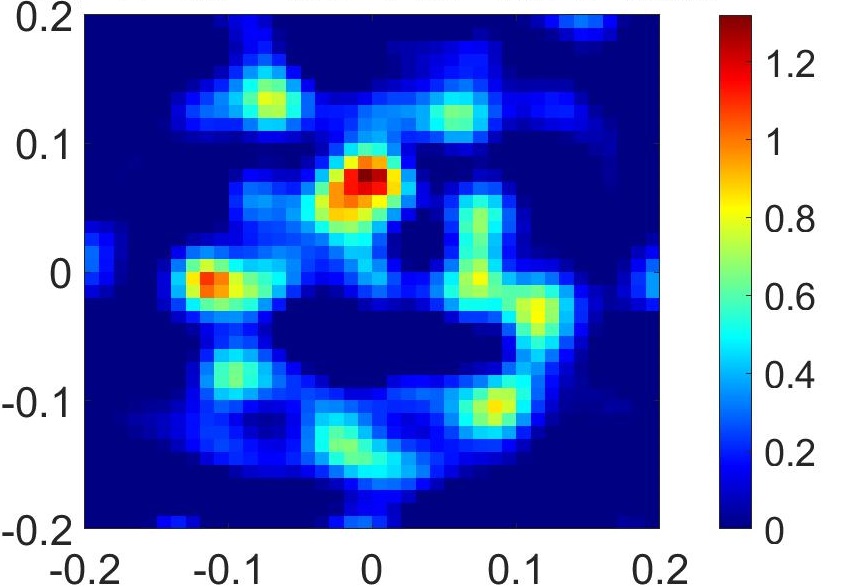}
		\subcaption*{PD-SOM $\operatorname*{Im}(\epsilon_r)$ \\(PSNR=0.8 dB, SSIM=0.61)}
	\end{subfigure}
	%%%%%%%%%%%%%%%%%%%%%%%%%%%%%%%%%%%
	\caption{Reconstruction of ``Austria" profile with 30\% noise in the measurements. (a) using DxPRIM, (b) using PD-SOM. x-axis and y-axis are in meters.}
	\label{Austria_noise} 
	\vspace{-0.5\baselineskip}	
\end{figure}

Next we increase the noise level to $30$\% in the measurements (for ``Austria" profile). The reconstruction results are shown in Fig. \ref{Austria_noise}, where Fig. \ref{Austria_noise}(a) is for DxPRIM and Fig. \ref{Austria_noise}(b) is for PD-SOM. It can be seen that for this level of noise, PD-SOM fails to provide good reconstruction, whereas, DxPRIM provides reconstructions with low but acceptable accuracy. The superior performance of DxPRIM in the presence of high noise can be attributed to its linear formulation compared to the non-linear formulation of PD-SOM. The linear techniques combined with the sparsity prior are more stable in the presence of noise whereas non-linear models can be highly sensitive to noise and small perturbations in measurements can lead to large error in the reconstruction results. %(error depends on how accurately the non-linear model represents the underlying true non-linear model). 

\begin{figure}[!h]
	\captionsetup[subfigure]{justification=centering}
	\centering
	%%%%%%%%%%%%%%%%%%%%%%%%%%%%%%%
	\adjustbox{minipage=1em,raise=6ex}{\subcaption{ }\label{}}% 
	\begin{subfigure}[t]{0.2\textwidth}
		\centering
		\includegraphics[width=1.2in]{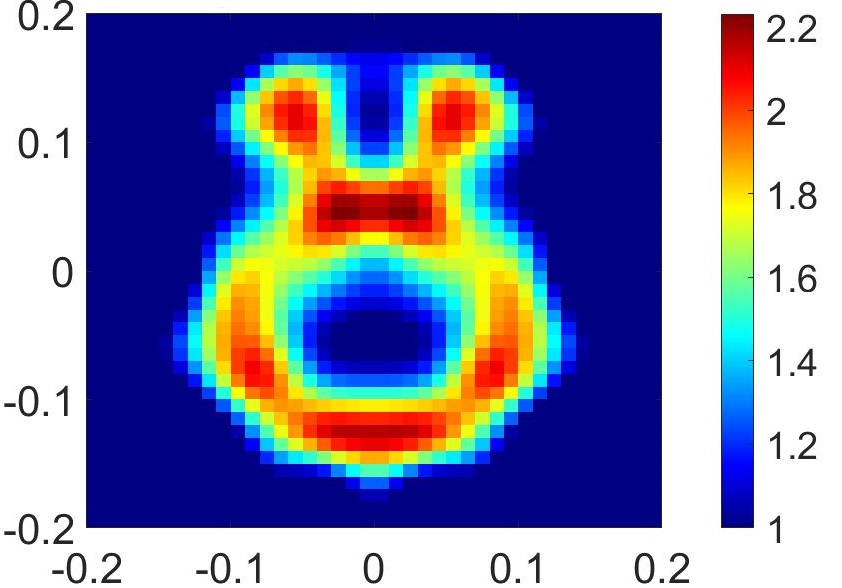}
		\subcaption*{DxPRIM $\operatorname*{Im}(\epsilon_r)$ \\(PSNR=12 dB, SSIM=0.80)}
	\end{subfigure}       %\hspace{3mm}
	\begin{subfigure}[t]{0.2\textwidth}
		\centering
		\includegraphics[width=1.2in]{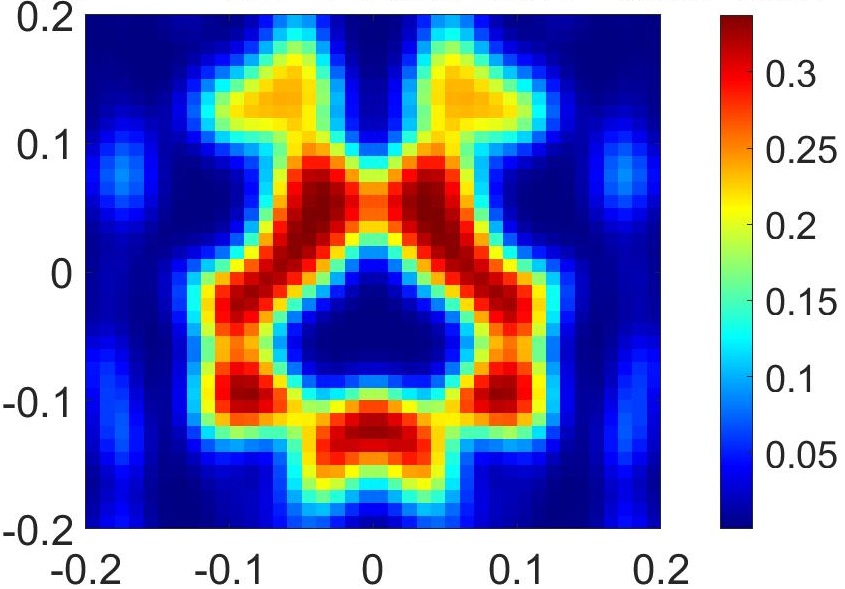}
		\subcaption*{DxPRIM $\operatorname*{Im}(\epsilon_r)$ \\(PSNR=9.5 dB, SSIM=0.77)}
	\end{subfigure} 
	\par\smallskip
	\adjustbox{minipage=1em,raise=6ex}{\subcaption{ }\label{}}%   
	\begin{subfigure}[t]{0.2\textwidth}
		\centering
		\includegraphics[width=1.2in]{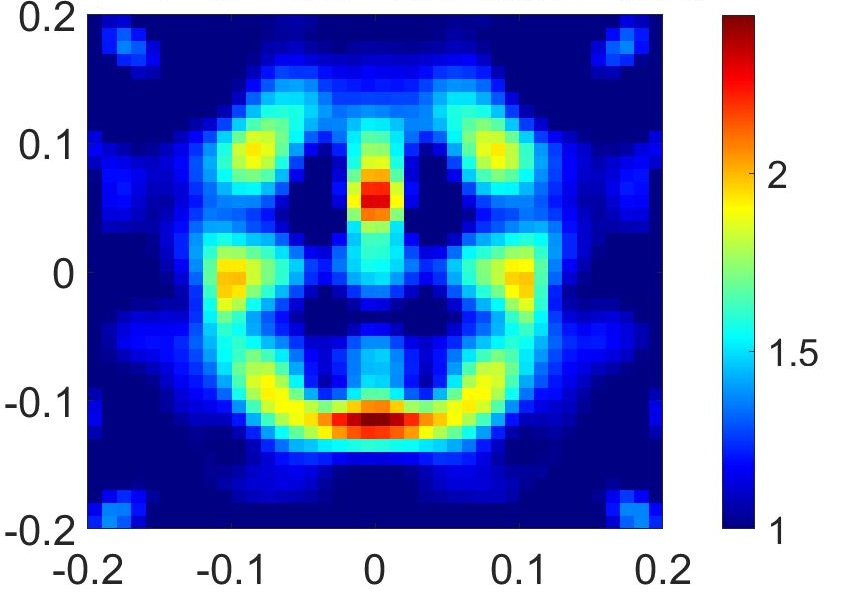}
		\subcaption*{PD-SOM $\operatorname*{Im}(\epsilon_r)$ \\(PSNR=3 dB, SSIM=0.62)}
	\end{subfigure}
	\begin{subfigure}[t]{0.2\textwidth}
		\centering
		\includegraphics[width=1.2in]{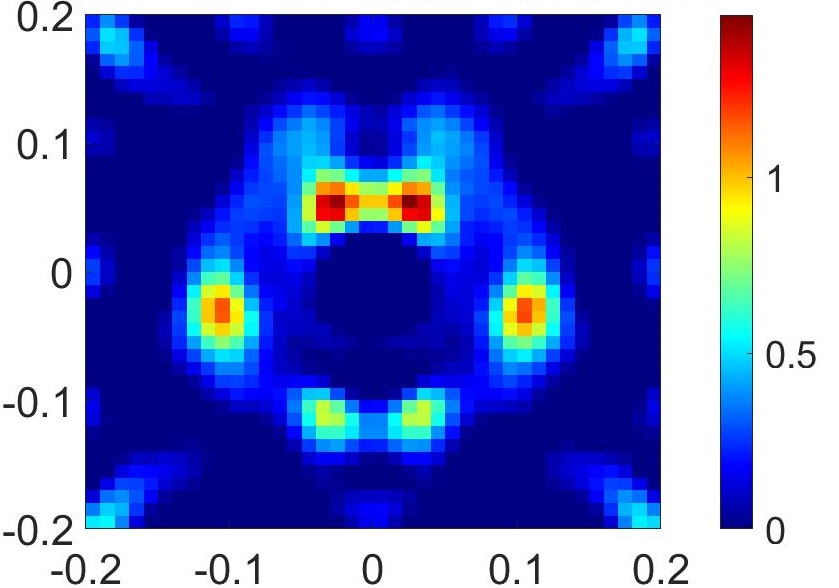}
		\subcaption*{PD-SOM $\operatorname*{Im}(\epsilon_r)$ \\(PSNR=1.5 dB, SSIM=0.54)}
	\end{subfigure}
	%%%%%%%%%%%%%%%%%%%%%%%%%%%%%%%%%%%
	\caption{Reconstruction of ``Austria" profile with $20$ transceivers, i.e., $M=20(20-1)/2 = 190$ measurements and $5$\% noise. (a) Reconstruction using DxPRIM, (b) Reconstruction using PD-SOM. x-axis and y-axis are in meters.}
	\label{Austria_few_nodes} 
	\vspace{-0.5\baselineskip}	
\end{figure}

In Fig. \ref{Austria_few_nodes}, we provide ``Austria" profile reconstruction results using only $20$ transceiver nodes or $M=M(M-1)/2 = 190$ measurements, which is one-fourth of the number of measurements used in the results shown in Fig. \ref{Austria}. This makes the inverse problem more ill-posed. We also add 5\% noise to the measurements. Fig. \ref{Austria_few_nodes} shows that DxPRIM is still able to recover the intricate ``Austria" profile, but with distortions (high distortion in $\operatorname*{Im}(\epsilon_r)$ profile). On the other hand, PD-SOM fails to recover the profile. Therefore, the linear DxPRIM approach is able to better tackle the ill-posedness compared to the non-linear PD-SOM method.

\subsection{Experimental Results}

\begin{figure}[!h]
	\captionsetup[subfigure]{justification=centering}
	\centering
	%%%%%%%%%%%%%%%%%%%%%%%%%%%%%%%
	\begin{subfigure}[t]{0.2\textwidth}
		\centering
		\includegraphics[width=1.3in]{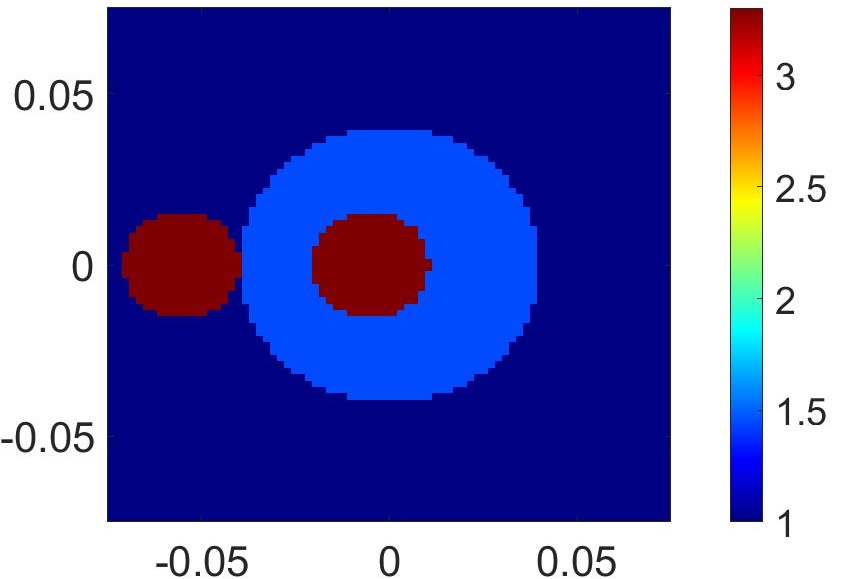}
		\subcaption{Ground Truth: $\operatorname*{Re}(\epsilon_r)$}
	\end{subfigure}  \medskip     \\%\hspace{3mm}\\
	\begin{subfigure}[t]{0.2\textwidth}
		\centering
		\includegraphics[width=1.3in]{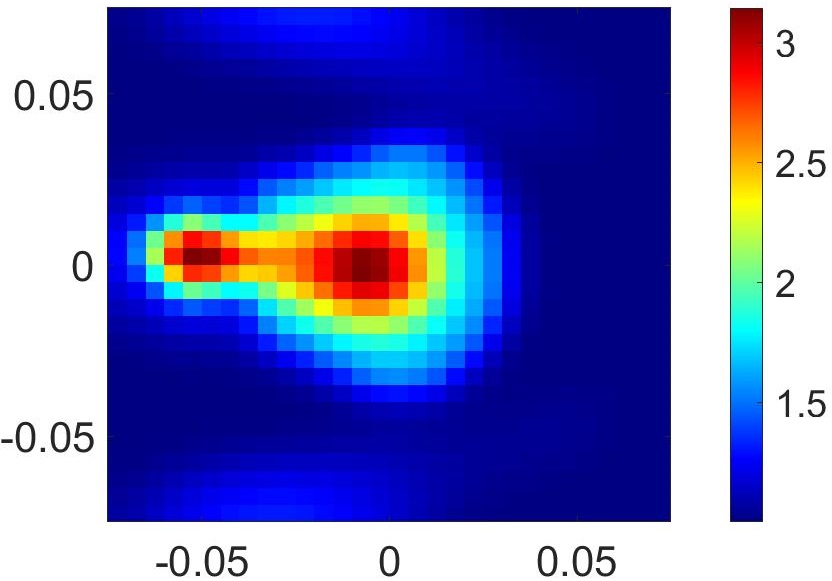}
		\subcaption{DxPRIM}
	\end{subfigure} \hspace{3mm}
	\begin{subfigure}[t]{0.2\textwidth}
		\centering
		\includegraphics[width=1.3in]{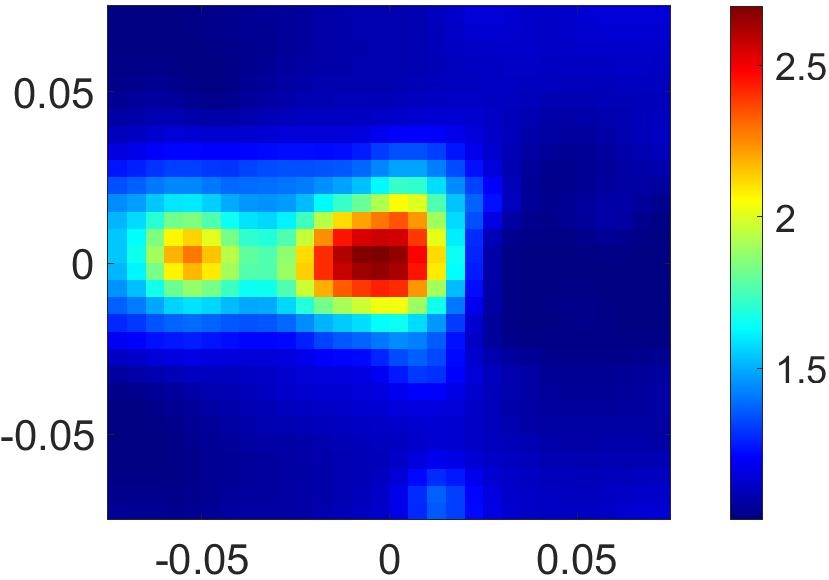}
		\subcaption{PD-SOM}
	\end{subfigure}
	%%%%%%%%%%%%%%%%%%%%%%%%%%%%%%%%%%%
	\caption{(a) ``FoamTwinDielTM" scatterer profile obtained from Fresnel Institute's experiments, (b) Reconstruction using DxPRIM (PSNR = 16 dB, SSIM = 0.84), (c) Reconstruction using PD-SOM (PSNR = 7 dB, SSIM = 0.69). x-axis and y-axis are in meters.}
	\label{Fresnel_inst}
	\vspace{-0.5\baselineskip}	
\end{figure}

To provide experimental results we use data provided by the Fresnel Institute, France where scattering data is collected in an anechoic chamber. The details of the experimental configurations have been previously described \cite{geffrin2005free}. We select the ``FoamTwinDielTM" dataset which uses a inhomogeneous scatterer profile as shown in Fig. \ref{Fresnel_inst}(a). It has overall size $15 \times 15$ cm$^2$ and consists of three cylindrical scatterers. The small red cylinder is of plastic material with $\epsilon_r=3 \pm 0.3$. The large blue cylinder is made up of foam with $\epsilon_r=1.45 \pm 0.15$. This large blue cylinder also contains an embedded small red cylinder with $\epsilon_r=3 \pm 0.3$. All these scatterers are considered lossless and the incident frequency is selected to be 2 GHz. 

Fig. \ref{Fresnel_inst}(b) and Fig. \ref{Fresnel_inst}(c) provide reconstruction results for ``FoamTwinDielTM" profile using DxPRIM and PD-SOM respectively. It can be seen that both methods are able to recover the scatterer information. However, our proposed DxPRIM technique provides better estimation of shape as well as permittivity.

\section{Conclusion}
In this paper we presented a new distorted wave extended phaseless Rytov iterative method to solve inverse scattering problems with phaseless data. It is based on a correction to the conventional Rytov approximation in strongly scattering, lossy media. The corrected Rytov approximation is then formulated into a distorted wave form. This distorted wave form is utilized in an iterative framework to achieve high quality reconstructions of strongly scattering, lossy objects. Using simulation and experimental reconstruction results for benchmark profiles (such as Austria profile), we show that our proposed technique, DxPRIM, outperforms a state-of-the-art PD-SOM technique.

\bibliographystyle{IEEEtran}
%\bibliography{strings,refs}
\bibliography{arXiv_Amar}

\end{document}